\documentclass[12pt,a4paper]{article}
\input epsf
\input rotate
\pdfoutput=1 
\usepackage{graphicx,subfig}
\usepackage{times}
\usepackage{tgtermes}
\usepackage{soul}
\usepackage{booktabs}
\usepackage{array}
\usepackage{multirow}
\usepackage{color}
\usepackage{collcell,xcolor}
\usepackage{colortbl}
\usepackage{ragged2e}
\usepackage{amsmath}
\PassOptionsToPackage{table,xcdraw}{xcolor}
\usepackage[table]{xcolor}
\definecolor{lightseagreen}{rgb}{0.125,0.7,0.67}
\definecolor{lightcoral}{rgb}{0.94,0.5,0.5}
\usepackage[utf8]{inputenc}
\newcommand{\HRule}{\rule{\linewidth}{0.5mm}}
\renewcommand{\thefootnote}{\fnsymbol{footnote}}

\begin{document}
\title{\bf Sensitivity of Indian summer monsoon rainfall forecast skill of CFSv2 model to initial conditions and 
        the role of model biases}

\author{K Rajendran$^{1,2,\footnotemark}$, Sajani Surendran$^{1,2}$ ,  \\
             Stella Jes Varghese$^{2}$  and Arindam Chakraborty$^{3}$ 
        \\
        \small $^{1}$ Multi-Scale Modelling Programme (MSMP), 
        \small        CSIR Fourth Paradigm Institute (CSIR-4PI), Bangalore, India \\
        \small $^{2}$ Academy of Scientific and Innovative Research (AcSIR), 
                      Ghaziabad, India \\
        \small $^{3}$ Indian Institute of Science (IISc), Bangalore, India \\
        } 
\date{\vspace{2in}\today}
\maketitle
        \vfill
        \noindent
        \HRule\\
\thefootnote{$^{*}$ Corresponding author address: Dr. K. Rajendran, 
                    Multi-Scale Modelling Programme (MSMP),
                    CSIR Fourth Paradigm Institute (CSIR-4PI), 
                    CSIR-NAL Belur, Wind Tunnel Road, Bangalore, 560 037. 
                    E-mail: rajend@csir4pi.in}
\newpage
\noindent
\baselineskip = 24pt
{\flushleft \bf \Large Abstract}

\noindent
This study analyses Indian summer monsoon (ISM) seasonal reforecasts by CFSv2 model, initiated from January 
(4-month lead time, L4) through May (0-month lead time, L0) initial conditions (ICs), to examine the cause 
for highest all-India ISM rainfall (ISMR) forecast skill with February (3-month lead time, L3) ICs.
The reported highest forecast skill for L3 ICs is based on correlation between
observed and predicted interannual variation (IAV) of ISMR. Other skill scores such as 
mean error, bias, RMSE, mean, standard deviation and coefficient of variation, 
indicate higher or comparable skill for April (L1)/May (L0) ICs.
Although theory suggests that skill degrades with increase in lead-time, CFSv2 shows 
highest skill with L3 ICs, due to predicting ISMR excess of 1983 for which other ICs fail. 
But, this correct forecasting  is caused by wrong forecast of 
La Ni\~{n}a, cooling of the equatorial central Pacific (NINO3.4) during the monsoon season, by L3 ICs.
In observation, near-normal sea surface temperatures (SSTs) prevailed over NINO3.4 and ISMR excess was due to variation of convection over equatorial Indian Ocean or EQUINOO which CFSv2 failed to capture with all ICs.  
Major results are reaffirmed by analysing an optimum number of experimental reforecasts by current version of 
CFSv2 initiated from five late-April/early-May ICs having shorter yet useful forecast lead time. 
These experimental reforecasts are found to have  least seasonal biases and
highest correlation skill score if 1983 is excluded.  
Model deficiencies such as over-sensitivity of ISMR to SST variation 
over NINO3.4 (ENSO) and unrealistic influence of ENSO 
on the EQUINOO, contribute to errors in ISMR forecasting. 
Whereas, in observation, ISMR is influenced by both ENSO and EQUINOO. 
The forecast skill for Boreal summer ENSO is found to be deficient in CFSv2 with
the skill being the lowest for L3/L4 ICs, hinting the possible influence of dynamical drift 
induced by long forecast lead-time. 
Rainfall occurrence despite strong cold bias over NINO3.4 in CFSv2, is associated with a stronger ocean-atmosphere coupling, with a shift of the SST-rainfall relationship pattern to slightly colder SSTs than the observed. 
These results warrant the need for minimisation of biases in SST boundary forcing to achieve improved ISMR forecasts.
\\

{\bf Keywords:} Indian Summer Monsoon Rainfall, Seasonal Reforecasts, Forecast Skill, Model Biases, ENSO, EQUINOO

\newpage
\section{Introduction}

    Rainfall received over India during the summer season (June to September, JJAS) is 
termed as the Indian summer monsoon rainfall (ISMR). There has been considerable year 
to year variation (known as interannual variation or IAV) in the quantum of ISMR that has a 
profound effect on the agricultural sector and the socioeconomic well-being of India. Hence, 
it is essential to predict ISMR or its departure in a season correctly to facilitate effective 
planning of agricultural and economic strategies, and water and hydel power management. 
Despite the challenges in modelling Indian summer monsoon due to its complex  features 
and multiple processes involved, coupled ocean-atmosphere general circulation models 
(CGCMs) have become an essential tool for dynamical seasonal prediction. The climate 
forecast system version 2 (CFSv2) model of the National Centers for Environmental 
Prediction (NCEP), USA, is an outcome of such efforts in 
recent years to improve dynamical prediction  and its forecast skill is widely studied (Saha et al. 2010, 
Krishnamurthy and Rai 2011, Pattanaik et al. 2012, Saha et al. 2014 etc.).
 Recently, this model is adopted 
by Ministry of Earth Sciences, Government of India, for dynamical seasonal prediction of 
Indian summer monsoon. 

    Over the tropics, the existence of slowly varying boundary conditions constitutes the 
basic premise of seasonal prediction (Charney and Shukla 1981). Anomalous IAV of sea surface temperature 
(SST) over the equatorial central Pacific associated with El Ni\~{n}o-Southern Oscillation 
(ENSO, Rasmusson and Carpenter 1983), is considered to be the primary source of predictability 
(Shukla and Wallace 1983). 
Krishnamurthy and Shukla (2011) examined the predictability of ISMR in eight CGCMs 
including CFS for forecast and predictability errors and estimated the doubling time of 
errors for rainfall over India, to be 4-14 days in the CFS against 4-7 days in other models. 
Forecast skill is to get better as the initial conditions (ICs) get closer to the prediction 
period and thus the highest forecast skill is expected for ICs with 0-month lead-time (L0). 
In other words, the skill is expected to increase (decrease) with decreasing (increasing) lead-time when 
considering the development of dynamical shift  in model with time (Slingo and Palmer 2011) and systematic 
biases caused due to deficient representation of physical processes in the model.   
Kumar et al. (2011) analysed CFS forecast skill of monthly mean SST and precipitation 
and showed that the skill rapidly decays with lead time. After a lead-time of about 
30-40 days, the forecast skill for monthly mean was found to deteriorate, with the SST 
anomalies in the tropical central/eastern Pacific playing a dominant role. Thus, for 
seasonal predictability, the conditions of the ocean state also become very important. 
They suggested the reduction in skill  is due to the large contribution from the 
atmospheric internal variability to monthly means. 

    Contrary to expectation and the understanding of significant ENSO spring predictability 
barrier and low predictability of ENSO forecasts during February-March (Webster and Yang 1992), 
CFSv2 predictions of ISMR with February ICs (3-month lead time, L3), are reported to 
have the maximum forecast skill (Pokhrel et al. 2016, Ramu et al. 2016, Pillai et al. 2018, Rao et al. 2019),
based on the correlation between observed and predicted IAV of ISMR during the analysis period. 
Further, the skill scores 
reported in previous studies vary considerably among themselves depending upon 
i) the region selected for averaging the summer season rainfall to estimate ISMR for 
each year, ii) the reference dataset used as observation and iii) the duration of the 
analysis period.  These seasonal forecast verifications are performed with datasets 
rarely exceeding 29 samples, which can also lead to highly uncertain scores (Schaeybroeck and Vannitsem 2019). 
However, an understanding of the impact of different ICs on ISMR forecast skill is 
fundamental and central to improving its predictability. Thus, it is imperative to 
understand what contributes to the forecast skill of February (L3) ICs. 
We focus on the factors which influence ISMR variability in CFSv2 by comparing its 
seasonal reforecasts (hindcasts) with observations/reanalyses, with emphasis on its 
dependence on SST boundary forcing. 
We analysed large datasets of 124 CFSv2 reforecasts initialised with ICs from 
1$^{st}$ January to 31$^{st}$ May, which are made available by NCEP. 
To reconfirm major results and to understand the advantage of choosing ICs which are 
nearer to the forecast period (JJAS) yet having useful lead-time,  we assessed the performance skill of the 
current version of CFSv2 by analysing its reforecasts initialised with an optimum subset of 5 
late-April/early-May ICs.

\section{Model, reforecasts, datasets and methodology}

	CFSv2 is a coupled dynamical forecast system with Global Forecast System model 
in triangular truncation of T126 ($\sim$0.9375$^{\circ}$ horizontal resolution) with 64 
hybrid sigma-pressure levels as the atmospheric component and Geophysical Fluid Dynamics 
Laboratory Modular Ocean Model v.4 (GFDL MOM4) with 0.25$^{\circ}$ horizontal 
resolution in equatorial region ($\pm$ 10$^{\circ}$ latitude and 0.5$^{\circ}$ 
elsewhere) as the ocean component (Saha et al. 2014).

        To examine the dependence of ISMR forecast skill of CFSv2 on ICs, we have analysed 
124 nine-month retrospective seasonal reforecasts or hindcasts (hereafter referred to as 
'CFSv2-NCEP') initiated from CFS Reanalysis based ICs on every 5$^{th}$ day starting 
from 1$^{st}$ January (4-month forecast lead time, 'L4') to 31$^{st}$ May (0-month 
lead time, 'L0'), with four reforecasts per day (at 00, 06, 12, 18 UTC) during 1982-2010 
period (Saha et al. 2010). These datasets are made available by NCEP in their web portal 
(https://www.ncdc.noaa.gov/data-access/model-data/model-datasets/climate-forecast-
system-version2-cfsv2). Thus, CFSv2-NCEP seasonal reforecasts initiated with 28 L4 ICs, 20 L3 ICs, 24 
L2 ICs, 24 L1 ICs and 28 L0 ICs, during 1982-2010, are analysed.
	
         We have carried out a set of nine-month reforecasts for the analysis period, with short 
lead times of late-April (L1) ICs (00UTC of 21 April and 26 April) and early-May (L0) ICs 
(00UTC of 1 May, 6 May and 11 May), using the current version of CFSv2 (which is being 
used for seasonal prediction as monsoon mission model by MoES, India), at the computing 
platform of Council for Scientific and Industrial Research (CSIR) Fourth Paradigm Institute, 
Bangalore (hereafter referred to as `CFSv2-CSIR'). 
Our analysis of retrospective forecasts revealed that the bias in representing the spatial distribution 
of climatological mean Indian summer monsoon rainfall is reduced with May ICs.  
The skill scores of correlation between the observed and predicted IAV of ISMR during 
1982-2010, are also found to be better for the ensemble means of April and May ICs, which are comparable 
to February ICs. This led us to choose an optimum subset of 5 late-April/early-May ICs which are 
close to the onset of the monsoon season yet having reasonable and useful lead time at the same time
yielding good skill scores in forecasting interannual variability of ISMR. For performing a set of 
experimental reforecasts which creates large outputs for about 3 decades, we have optimised the 
number of ICs to 5 which can yield the best skill score, i.e., two late-April and three early May ICs. 
These runs are analysed to verify major results on forecast skill for ISMR and summer-time 
ENSO, in the current version of the model. 
In addition, for the special year of 1983, we have carried out CFSv2-CSIR reforecasts with 
February (L3) ICs as well.

    For validation, 0.25$^{\circ} \times$ 0.25$^{\circ}$ gridded India Meteorological Department 
(IMD) rainfall (Pai et al. 2014), Global Precipitation Climatology Project (GPCP) version 2.3 data (Adler et al. 2003) 
and Hadley Centre Ice and SST (HadISST) data (Rayner et al. 2003) are used. Daily optimum interpolation 
SST version 2.1 (OISSTv2.1 at 0.25$^{\circ} \times$ 0.25$^{\circ}$ horizontal resolution)
data (Reynolds et al. 2007) is also analysed. For validation of 850 hPa winds, we use 5$^{th}$ generation 
European Centre for Medium Range Weather Forecast Reanalysis (ERA5) data (Hersbach et al. 2020).
NINO3.4 (170$^{\circ}$-120$^{\circ}$W; 5$^{\circ}$S-5$^{\circ}$N) SST anomaly (normalised 
by standard deviation) is used as ENSO index. ENSO index $>$ 1 ($<$1) indicates El Ni\~{n}o 
(La Ni\~{n}a).  It is important to note that our analysis focuses on ENSO during the Indian summer 
monsoon season (i.e. JJAS). Negative of the anomaly of surface zonal wind at the equatorial 
Indian Ocean (IO, 60$^{\circ}$-90$^{\circ}$E; 2.5$^{\circ}$S-2.5$^{\circ}$N) estimated from 
ERA5 is used as the index for equatorial Indian Ocean oscillation (EQUINOO, Gadgil et al. 2004).

    For estimating seasonal mean ISMR for each year, we use the rainfall averaged over the 
monsoon region (Gadgil et al. 2019). Anomalies of ISMR, and indices of ENSO and EQUINOO are 
standardised with their standard deviation. For assessing the performance of forecasting 
the IAV of ISMR, basic skill scores such as ISMR temporal mean, standard deviation and 
coefficient of variation (CV) are used. Deterministic skill scores such as mean error, bias and 
root-mean-square error (RMSE) are also used. Details of 
these methods are provided in Appendix 1. To assess climatological mean monsoon rainfall pattern over India, 
statistics with respect to IMD rainfall viz. spatial pattern correlation coefficient (PCC), 
ratio of standard deviation against observed (SD) and climatological bias as percentage 
of observed (bias), are computed.

\section{Dependence of ISMR forecast skill  on initial conditions}

	Figure 1a shows the interannual variation (IAV) of standardised ISMR anomalies from 
IMD observation and deterministic ensemble mean of CFSv2-NCEP L3 reforecasts.
The correlations for deterministic reforecasts against the observed for 1982-2010 period ($\gamma$) 
for CFSv2-NCEP reforecasts with L0 to L4 ICs are written in Fig. 1a.
It can be seen that the ISMR forecast skill based on correlation 
is the highest for CFSv2-NCEP L3 ($\gamma$=0.44) followed 
by CFSv2-NCEP April (L1) ICs ($\gamma$=0.35).
We can see that the performance of ensemble mean of CFSv2-CSIR reforecasts with 
late-April/early-May ICs (Fig. 1a) is comparable with that of L3 ($\gamma$=0.38). 

Corresponding IAV of Boreal summer ENSO index (NINO3.4 SST) anomalies is shown in Fig. 1b.
ISMR excess of 1983 in L3 is associated with erroneous Boreal summer (JJAS)
La Ni\~{n}a forecast when the observed SST condition was neutral over NINO3.4 (Fig. 1b). 
The extreme ISMR departure in 1983 is captured only by February ICs, in magnitude and sign. 
To some extent, the departure of 1994 is also captured by CFSv2-NCEP L3. However, 
all ICs fail to forecast the departures of 1985, 1990, 1997, 1998 and 2006. 
For these special years, departures of ISMR and ENSO index predicted by the 
ensemble means of CFSv2-NCEP reforecasts with L4 to L0 ICs are compared in 
Table 1. We can see that the ISMR departures in CFSv2-NCEP are largely influenced by 
the sign and magnitude of their ENSO index forecasts; ISMR deficits are associated with 
El Ni\~{n}o or the anomalous warming of NINO3.4 SST and excesses are associated with La Ni\~{n}a or the
anomalous cooling of SSTs over NINO3.4 region. The inverse relationship and interaction between 
ENSO and ISMR during boreal summer, are well-documented (Walker and Bliss 1932, Walker 1933, Sikka 1980, Torrence and Webster 1999 etc.). This relationship is modulated on decadal timescales 
(Kumar et al. 1999, Chen et al. 2010, Kumar et al. 2006, Azad and Rajeevan 2016, Fan et al. 2017). Most importantly, 
in CFSv2-NCEP reforecasts, ISMR is shown to be having over-sensitivity to ENSO, especially to the 
SST fluctuation over the equatorial central Pacific region (Vishnu et al. 2019).

    Further examination of yearly ISMR departures, reveals that skill is not better in 
CFSv2-NCEP L3 compared to CFSv2-CSIR (Fig. 1a), though their correlations are 
0.44 and 0.38 respectively. The difference between these two correlations is significant only 
at 0.73\% confidence level. However, the previous studies 
(Pokhrel et al. 2016, Ramu et al. 2016, Pillai et al. 2018, Rao et al. 2019) have all reported the skill 
improvement in CFSv2 with February ICs with such differences in correlations, though the exact 
correlation values vary from one study to another depending upon the region selected 
to compute average seasonal ISMR, the data used as reference/observation and the 
duration of the analysis period. 

    In observation, the ISMR departures of 1983, 1994, 1985, 1990, 1997, 1998 and 2006 
are not strongly (and inversely) related to ENSO anomalies (Table 1). The excesses of 1983 
and 1990 are associated with neutral ENSO phases and excesses of 1994 and 2006 are 
associated with mild warming and deficit of 1985 is associated with strong cooling over 
NINO3.4. In spite of strong El Ni\~{n}o in 1997 and La Ni\~{n}a in 1998, the ISMR 
remained close to normal in observation. In comparison, ISMR departures of 
1985, 1990, 1997, 1998 and 2006, are associated with very intense NINO3.4 SST 
anomalies in the model and all ICs forecast the inverse ISMR departures. During these 
years the inverse relationship is strong and evident in the model and the dominant 
driving force determining the ISMR departure remains to be ENSO with all ICs. 
Compared to observation, the model tends to show amplified  ENSO anomalies 
(skewed for cold events), more for earlier ICs of L4-L2 and most conspicuously for L3. 
For 1983 and 1994, larger errors are seen in ENSO predictions, with the largest error 
amplitudes for L3. 

    It can be seen that the correlation for CFSv2-NCEP L3 falls to 0.4 which is lower than 
the corresponding score ($\gamma$=0.42) for CFSv2-CSIR, if we exclude 1983 (Fig. 2). 
So the improved ISMR forecast skill of L3 is contributed by its prediction of 1983 ISMR excess. 
Next, we applied other deterministic verification scores such as the mean error, bias, and RMSE 
during the analysis period, for assessing the forecast skill of CFSv2-NCEP L3 and
CFSv2-CSIR (Table 2). These skill scores are clearly improved in CFSv2-CSIR reforecasts
compared to CFSv2-NCEP L3 (Table 2). The deficiencies of underestimation of the mean 
(dry bias) and standard deviation (reduced variability) of ISMR, also get improved and 
CV becomes the closest to the observed in CFSv2-CSIR.

    Model intercomparison studies in the past had suggested that models which are skilful in 
representing climatological mean summer monsoon rainfall are more adept in simulating IAV 
of ISMR (Sperber and Palmer 1996, Gadgil and Sajani 1998). 
CFSv2 is found to have reasonable skill in capturing the spatial distribution 
of climatological JJAS mean rainfall, SST and 850 hPa winds over Indian region.  The mean bias 
is lower in CFSv2-CSIR than in L3 (Fig. 3). Still, there exists underestimation of rainfall (dry bias) 
over central India coinciding the seasonal monsoon trough zone, and uniform wet bias and 
wide-spread underestimation of SST (cold bias) over the Indian Ocean and West Pacific. 
Oceanic regions with enhanced rainfall are associated with convergence and colder SSTs. 
Wet bias over equatorial IO has zonal asymmetry with the maximum over the eastern equatorial 
Indian Ocean (EEIO) with strong westerly wind biases and low-level convergence. 
Cold bias over the equatorial central Pacific is found to be associated with the strengthening of ITCZ  
(wet bias) and with cyclonic wind bias slightly north of equator (not shown).

    Dry bias over India is larger in CFSv2-NCEP L3 compared to CFSv2-CSIR. The pattern 
correlation coefficient (PCC), standard deviation (SD) and mean bias are largely comparable among 
ensemble means of CFSv2-NCEP and CFSv2-CSIR reforecasts (Table 3). But,  the PCC is 
slightly larger for reforecasts with May ICs. Similarly, standard deviation and bias are clearly 
improved in CFSv2-CSIR reforecasts with late-April/early-May ICs and it is better than L3 in 
representing mean monsoon rainfall over the Indian region. This is expected as atmospheric and oceanic 
states are close to JJAS. The increase in bias as lead-time increases, indicates the role of 
dynamical drift in the model.

{\flushleft \bf 3.1 ISMR-ENSO relationship}

    The leading factor determining IAV of ISMR is the strong relationship between ISMR and 
ENSO in which there is an increased propensity of droughts during El Ni\~{n}o and of excess 
rainfall during La Ni\~{n}a (Sikka 1980). It can be gleaned from Fig. 1 that 8 out 12 excess events 
are associated with La Ni\~{n}a and 8 out of 12 deficit
events are associated with El Ni\~{n}o in CFSv2-NCEP L3. There are no large excess 
(large deficit) associated with El Ni\~{n}o (La Ni\~{n}a). All large excesses 
(large deficits) are associated with La Ni\~{n}a (El Ni\~{n}o). Thus, ISMR-ENSO relationship  
is much stronger in CFSv2-NCEP L3 with a correlation of -0.85 than in observation 
($\gamma$=-0.44) where other factors do influence ISMR (Fig. 4). The strongest correlation 
is seen for L4 followed by L3  and the correlation is the lowest  for L0 ICs. For CFSv2-CSIR 
reforecasts with late-April/early-May ICs, the correlation is -0.79 which is closer to the 
observation compared to L3. It is to be recalled that its ISMR forecast skill is also comparable 
with L3 for 1982-2010 period which becomes better ($\gamma$=0.42) than that of L3 
($\gamma$=0.40) when 1983 is excluded from the analysis period (Fig. 2). Thus, the 
correct forecast of 1983 ISMR excess as a result of an erroneous La Ni\~{n}a forecast 
by L3  contributed to the seemingly higher IAV correlation for L3. But, other skill scores 
do not show higher ISMR forecast skill for L3 (Table 2). Moreover, the Boreal summer 
ENSO forecast skill is the lowest for L3 (Fig. 1b).  This makes it necessary to analyse its ENSO 
forecast skill during Boreal summer, in detail.

\vspace{1cm}

{\flushleft \it Boreal summer ENSO forecast skill}

    CFSv2-NCEP L3 appears to have serious deficiency in forecasting summer-time ENSO
(Fig. 1b). The forecast skill for JJAS ENSO index is found to be the lowest in CFSv2-NCEP L3 
and L4 ($\gamma$=0.59) compared to those in L2 to L0, and in CFSv2-CSIR 
($\gamma$=0.76). We have seen that the skill is much higher when all 124 reforecasts 
are considered together with the correlation of their median with the observed being 0.74 (not shown). 
The verification of performance of CFSv2 in predicting the warm and cold SST anomalies over the 
NINO3.4 region can be done from the classification of hits, misses and false alarms in CFSv2-NCSP L3 and CFSv2-CSIR 
reforecasts (Table 4). The forecasts miss several events and there are a few false alarms as well. 
The number of misses and false alarms for cold and warm events is more for L3 forecasts.
Thus, the performance is slightly better for CFSv2-CSIR reforecasts with late-April/early-May ICs. 

Monthly forecast skill scores for ENSO indices against the observed during the 
analysis period (written in Figs. 5a-d) clearly manifest the bias in L3 ENSO forecasts for 
June, July, August and September. During 1983, L3 predicted neutral condition in June 
and thereafter strong La Ni\~{n}a which kept intensifying from July to September. 
In contrast, in observation, NINO3.4 was having El Ni\~{n}o in June, neutral conditions 
in July and August, and a mild cold anomaly in September. The forecast skill for 1982-2010 
period, systematically drops from June to September with the least skill exhibited in September (Fig. 5).  

Correspondingly, the relationship of NINO3.4 SST with local NINO3.4 rainfall and remote impact on 
ISMR in 1983, show model biases in L3. In observation, there is enhanced NINO3.4 rainfall 
associated with El Ni\~{n}o in June which tends to become normal as SSTs approach climatology 
and then develops  to a cold anomaly by September. Accordingly, ISMR varies from below normal in 
June to normal in July to  large excesses in August and September. This is consistent with the 
inverse relationship between ISMR and ENSO. In L3, ENSO condition is near-neutral with excess 
rainfall over NINO3.4 in June which drops to strong La Ni\~{n}a in July which intensifies thereafter 
with deficit rainfall over NINO3.4. This results in above-normal ISMR in June and large excesses 
in July to September of 1983.  

The strong association of local rainfall with NINO3.4 SST, even with  cold bias over the equatorial 
Pacific Ocean in CFSv2 (not shown), can be understood from the SST-rainfall relationship over 
NINO3.4 from June to September of 1982-2010 (Fig. 6). 
Figure 6 shows the number of points for each 0.25$^{\circ}$C SST and 0.5 mm/day rainfall 
bin, along with the variation of mean rainfall with SST. The observed relationship 
($\gamma$=0.59) shows that the rainfall along with the mean steadily increases with SST 
from about 27$^{\circ}$C with high propensity of rainfall for SSTs above this threshold. 
In CFSv2-NCEP L3, there is a slight shift in the SST-rainfall relationship towards colder 
SSTs (Fig. 6) with the number of observations above 28$^{\circ}$C becoming much lower 
than the observed. This is consistent with the finding that the SST-rainfall pattern in coupled 
models is similar to the corresponding observation or atmosphere-only version, except for a 
shift of the pattern to colder/warmer SSTs as per their seasonal mean cold/warm bias (Rajendran et al. 2012). 

\section{Role of bias in SST boundary forcing} 

In CFSv2, the SST-rainfall association/relationship over NINO3.4 is stronger with a correlation 
of 0.65 than observed (Fig. 6), which in turn seems to have a remote impact on ISMR. Thus the 
ISMR prediction depends highly on ENSO. This indirectly implies that the reduction of SST bias 
over central Pacific can contribute to improvement in ISMR forecast skill.
Further, daily SST averaged over the NINO3.4 region shows that SST starts falling sharply after 
the beginning of monsoon season in 1983 (Fig. 7). The dropping of SST is steep and large. 
The characteristics of the evolution of 1983 SST over NINO3.4 for L3 ICs remain the same in the 
current version of CFSv2 as well (i.e., CFSv2-CSIR initiated with February ICs). The build-up 
of bias hints at the role of dynamic drift and model bias resulting in colder SSTs by the 
summer months for L3.  Given the high sensitivity of ISMR to NINO3.4 SST boundary forcing, 
systematic approach to minimise SST bias is essential to achieve the potential predictability.

{\flushleft \bf 4.1 Link of ISMR with Equatorial Indian Ocean} 

Another mode of SST variability in the equatorial Indian Ocean (IO), is the occurrence of opposite 
SST anomalies over eastern equatorial IO (EEIO, 90$^{\circ}$-110$^{\circ}$E; 
10$^{\circ}$S-0$^{\circ}$) and western equatorial IO (WEIO, 50$^{\circ}$-70$^{\circ}$E; 
10$^{\circ}$S-10$^{\circ}$N), known as the Indian Ocean dipole (IOD, Saji et al. 1999). 
Climatologically, IO is warmer in the east supporting more convection than in the west 
during monsoon. Positive IOD phase is characterized by weakening or reversal of 
climatological zonal SST gradient with suppression (enhancement) of convection over 
east (west) and anomalous winds blow from east to west along the equator, lifting up 
of thermocline and mixed layer of the east. However, the relationship between ISMR 
and IOD during JJAS is found to be rather weak, with the correlation coefficient not 
significantly different from zero, and only about 1\% of ISMR variance explained by IOD (Sajani et al. 2015).

The atmospheric counterpart of IOD, the equatorial Indian Ocean Oscillation (EQUINOO), 
with its positive (negative) phase associated with enhanced convection over WEIO (EEIO) 
and suppressed convection over EEIO (WEIO) is found to play an important role in 
determining IAV of ISMR (Gadgil et al. 2004) with the positive (negative) phase favourable 
(unfavourable) for ISMR. As the positive (negative) EQUINOO phase is associated 
with an easterly (westerly) anomaly of the zonal wind over the central equatorial IO, 
the EQUINOO index is based on the surface zonal wind anomaly over this region. 
Although EQUINOO is considered to be the atmospheric component of the coupled 
IOD mode, unlike ENSO, they are not as tightly coupled (Sajani et al. 2015) with correlation 
between their indices being only $\sim$0.45. 

Spatial distribution of 1983 anomalies of rainfall, SST and winds show that in observation, 
the enhanced rainfall over India is due to a positive EQUINOO (convective WEIO), 
whereas, in CFSv2-NCEP L3, it is associated with La Ni\~na (Figs. 8a and b). 
In CFSv2-CSIR, ISMR deficit is associated with El Ni\~no (Fig. 8c). Both forecasts do not show 
the important positive association between monsoon rainfall over WEIO and Indian region. 
Further analysis reveals that in CFSv2, EQUINOO appears to occur 
due to the impact of ENSO on the equatorial Indian Ocean. The impact of El Ni\~{n}o results in 
warming and enhancement of rainfall over WEIO and cooling and suppression 
of rainfall over EEIO region extending up to the West Pacific and over the Indian region. 
During La Ni\~{n}a opposite impacts occur over WEIO, EEIO and Indian region. 
Thus, ENSO elicits an inverse relationship between WEIO and EEIO which is analogous to the 
EQUINOO characteristics over the Indian Ocean. At the same time, ENSO impact 
results in an inverse relationship between ISMR and WEIO rainfall which is in contrast to 
their observed relationship associated with EQUINOO. 
The fact that enhanced cross-equatorial flow is associated with enhanced 
off-equatorial diabatic heating along the monsoon convergence zone over India results 
in a strong positive correlation between convection over WEIO and ISMR which is observed during
strong EQUINOO events. But, this relationship fails to exist in CFSv2. In contrast,
`EQUINOO-like' events with opposite poles of anomalies of SST, rainfall and circulation 
over equatorial WEIO and EEIO occur which are induced by ENSO in the model (not shown). 
This endorses opposite relationship between WEIO convection and ISMR. 

In CFSv2, ENSO and EQUINOO result in reinforcing each other's inverse impact on ISMR,
in contrast to observation where they tend to oppose each other (Vishnu et al. 2019). 
This leads to much stronger than the observed 
inverse relationship between ENSO and ISMR in CFSv2 (Fig. 4). 
Correspondingly, ENSO and EQUINOO show an intense
correlation between them; 0.83 in CFSv2-NCEP L3 
(largest among ICs) and 0.58 in CFSv2-CSIR. In contrast, in observation they are almost 
independent ($\gamma$=0.14, Fig. 9a). Resultantly, ISMR-EQUINOO relationship is also too strong in 
CFSv2-NCEP L3 ($\gamma$=-0.77, the largest among ICs, $\gamma$=-0.56 for 
CFSv2-CSIR) which is opposite to the observed ($\gamma$=0.54) relationship (Fig. 9b). 

Forecast of 1994 ISMR departure by L3 was also due to an erroneous La Ni\~{n}a forecast 
when in reality ISMR was excess only due to positive EQUINOO (coloured as green for positive 
EQUINOO and red for negative EQUINOO events in Table 1). In CFSv2, ISMR departure 
is almost entirely decided by ENSO whereas in observation EQUINOO is found to play a decisive 
role in several years (Table 1). It can be seen that excesses of 1983 (with neutral ENSO condition) and 
1994 and 2006 (with mild warm ENSO anomalies)  are due to positive EQUINOO events. 
ISMR of 1985 was below normal due to negative EQUINOO despite having very 
strong cold ENSO anomaly. Normal monsoons of 1997  and 1998 are due to positive 
and negative EQUINOO events despite having very strong El Ni\~{n}o and La Ni\~{n}a 
respectively. Inability of CFSv2 to forecast EQUINOO events independent of ENSO, 
made it impossible for the model to forecast ISMR anomalies of 1985, 1990, 1997, 1998 
and 2006 with almost all ICs (Figs. 1a and Table 1).

\section{Concluding Remarks}

This study attempts to understand what contributes to the highest ISMR forecast skill for 
CFSv2 February (3-month forecast lead time, L3) ICs as reported in previous studies. We analysed 124 retrospective 
nine-month reforecasts by CFSv2 with January (4-month forecast lead time, L4) through May (0-month forecast lead time, L0) ICs, provided by NCEP for 1982-2010 period (referred to as CFSv2-NCEP reforecasts). Our analysis reveals that the reported higher forecast skill for 
February (L3) ICs was based on a single skill score of correlation between observed 
and predicted ISMR departures during the analysis period. In contrast, other skill scores 
such as the mean error, interannual bias and RMSE, and the mean, standard deviation and coefficient of variation, 
indicate higher or comparable forecast skill for April/May (L1/L0) ICs. 
Climatological bias in mean summer monsoon rainfall over India 
is also found to be the least with L1/L0 ICs. 
These results are reconfirmed through the analysis of a set of experimental reforecasts by the current version of CFSv2 
with an optimum subset of 5 late-April/early-May ICs which are having shorter yet useful
forecast lead times (referred to as CFSv2-CSIR reforecasts). Correspondingly, reforecasts with 
late-April/early-May ICs yield a correlation skill score comparable to that of L3 and 
the deterministic ISMR forecast skill is found to be the best with late-April/early-May ICs 
for 1982-2010 period, if 1983 is excluded. 

The success of CFSv2-NCEP L3 in forecasting a 
single event, i.e., excess ISMR departure in 1983, contribute to its higher IAV 
correlation of 0.44. The correlation is 0.38 for CFSv2-CSIR late-April/early-May ICs 
which is significantly different from that of CFSv2-NCEP L3  with only 73\% confidence. These
correlations become 0.40 and 0.42 for CFSv2-NCEP L3 and CFSv2-CSIR respectively, if 1983 
is excluded from the analysis period of 1982-2010. 
Further, we find that the success of CFSv2-NCEP L3 in forecasting 1983 ISMR 
excess is due to its wrong forecast of La Ni\~{n}a 
(unlike L1 and L0 ICs) during Boreal summer of 1983. 
Our analysis thus suggests the importance of initialising seasonal forecasts from April/May ICs.

CFSv2's common deficiencies such as the over-intensified influence of ENSO on ISMR and on
variation of SST, rainfall and circulation over the equatorial Indian Ocean, are also important factors which contribute
to errors in ISMR forecasting. 
In CFSv2, ISMR is almost entirely decided by ENSO related SST boundary forcing, 
with no link between variabilities of ISMR and convection over equatorial Indian Ocean associated with EQUINOO. 
In contrast, in observation, ISMR is influenced by both ENSO and EQUINOO independently. 

Central Pacific was under the sway of  El Ni\~{n}o till June 1983. All forecasts were 
initiated when El Ni\~{n}o was prevailing with active convection over NINO3.4. 
CFSv2 is known to develop pronounced wet and cold bias over the central Pacific. 
The fact that CFSv2-NCEP L3 with long lead-time ended in forecasting La Ni\~{n}a in summer
hints at the possible role of wet bias and associated winds resulting in stronger cooling of NINO3.4 
ocean surface for L3. This also implies that the persistence of errors in atmospheric 
circulation due to imperfections in physical processes could eventually lead to 
large-scale bias in ocean circulation and surface temperatures. This can be manifested 
in larger magnitudes in forecasts with longer lead-times. Improvements in 
atmospheric model physics schemes and experiments with observed SST forced 
atmosphere-only component of CFSv2 can throw further light on these aspects. 
It is also important to see if the ocean model of CFSv2 can simulate oceanic modes 
correctly when forced with realistic atmospheric circulation and fluxes. Our analysis 
suggests the need for a systematic approach to minimise the biases in SST boundary 
forcing in CFSv2, to achieve improved ISMR forecasts.

\thispagestyle{empty}
{\flushleft \bf Acknowledgements}

     We thank Prof. JS for many useful discussions and suggestions. This study is supported by 
National Monsoon Mission Phase-II project (GAP-1013) funded by the Ministry of Earth Sciences 
(MoES), Government of India. We acknowledge the availability of CSIR-4PI HPC, Anantha.
The authors acknowledge valuable discussions with Prof. Ravi S Nanjundiah of IISc 
on the implementation of CFSv2 on Anantha. We are thankful to the NOAA National Centers for 
Environmental Information and National Centers for Environmental Prediction (NCEP) 
which is responsible for Climate Forecast System (CFS), for making available 
the CFSv2 reforecasts' outputs (Saha et al. 2014) on their web portal  
(https://www.ncdc.noaa.gov/data-access/model-data/model-datasets/climate-forecast-system-version2
-cfsv2). We acknowledge the NCEP Climate Forecast System 
Reanalysis initial conditions (Saha et al. 2010) provided in the above portal which are used for carrying 
out seasonal reforecasts of CFSv2 with late-April/early-May initial conditions during 1982-2010. 
We also acknowledge the India  Meteorological Department (IMD) for providing 0.25$^{\circ} 
\times$ 0.25$^{\circ}$ gridded rainfall (http://imdpune.gov.in/Clim\_Pred\_LRF\_New/
Grided\_Data\_Download.html, Pai et al. 2014), Global Precipitation Climatology Project (GPCP) version 2.3 data 
(https://psl.noaa.gov/ data/gridded/data.gpcp.html, Adler et al. 2003), Hadley Centre Ice and SST 
(HadISST) data (http://badc. nerc.ac.uk/view/badc.nerc.ac.uk\_\_ATOM\_\_dataent\_hadisst, Rayner et al. 2003),
5$^{th}$ generation European Centre for Medium Range Weather Forecast Reanalysis 
(ERA5) data (https://www.ecmwf.int/en/ forecasts/datasets/reanalysis-datasets/era5, Hersbach et al. 2020) and
NOAA High Resolution SST data  provided by the NOAA/ OAR/ESRL PSL, Boulder, 
Colorado, USA, at their Web site at https://psl. noaa.gov/ (Reynolds et al. 2007). 

\newpage
{\flushleft \bf References}

\begin{enumerate}

\item {Adler, R. F., Huffman, G. J. Chang, A. and co-authors, The version 2 {Global Precipitation 
         Climatology Project (GPCP)} monthly precipitation analysis (1979-present). 
         2003, {\it J. Hydrmeteorol.}, {\bf 4}, 1147-1167.}
         
\item {Azad, S., and Rajeevan, M., Possible shift in the ENSO-Indian monsoon rainfall 
         relationship under future global warming. 2016, {\it Sci. Rep.}, {\bf 6(1)}, 20145.}

\item {Charney, J. G. and Shukla, J., Predictability of monsoons. 1981, 
         {\it Monsoon dynamics}, {\bf 4}, 99-109.} 
         
 \item {Chen, W., Dong, B., and Lu, R., Impact of the Atlantic Ocean on the multidecadal 
          fluctuation of El Ni\~no-Southern Oscillation-South Asian monsoon relationship in a 
          Coupled General Circulation Model. 2010, {\it J. Geophys. Res.}, {\bf 115}, D17109.} 
            
\item {Fan, F., and co-authors, Revisiting the relationship between the south Asian summer 
         monsoon drought and El Ni\~no warming pattern. 2017, {\it Atmos. Sci. Lett.}, 
         {\bf 18(4)}, 175-182.}
         
 \item {Gadgil, Sulochana, and Sajani, Surendran,  Monsoon precipitation in the AMIP runs. 
         1998, {\it Clim. Dyn.}, {\bf 14}, 659-689.}
         
\item {Gadgil, Sulochana., Vinayachandran, P. N., Francis, P. A., and Gadgil, S., Extremes 
         of Indian summer monsoon rainfall, ENSO, equatorial Indian Ocean oscillation. 
         2004, {\it Geophys. Res. Lett.}, {\bf 31}, L12213.}
      
\item {Gadgil, Sulochana., Rajendran, K., and Pai, D. S., A new rain-based index for the 
        Indian summer monsoon rainfall. 2019, {\it Mausam}, {\bf 70(30)}, 485-500.}
         
\item {Hersbach, H., Bell, B., Barrisford, P., Hirahara, S., and co-authors,  
         The ERA5 global reanalysis. 2020,  {\it Quart. J. Roy. Meteorol. Soc.}, 
         {\bf 146 (730)}, 1999-2049.} 

\item {Krishnamurthy, V. and Rai, S., Predictability of South Asian monsoon circulation in the 
         NCEP Climate Forecast System. 2011, {\it Adv. Geosci.}, {\bf 22}, 65-76.}
                  
 \item {Krishnamurthy, V. and Shukla, J., Predictability of the Indian monsoon in coupled 
         general circulation models.  2011, {\it COLA Tech. Rep.}, {\bf 313}.}     
            
\item {Kumar, A., Chen, M. and Wang, W., An analysis of prediction skill of monthly mean 
         climate variability. 2011, {\it Clim. Dyn.}, {\bf 37}, 1119-1131.} 
           
 \item {Kumar, K. K., Rajagopalan, B., and Cane, M. A., On the weakening relationship between 
          the Indian monsoon and ENSO. 1999, {\it Science}, {\bf 284(5423)}, 2156-2159.}
          
\item {Kumar, K. K., Rajagopalan, B., Hoerling, M., Bates, G., and Cane, M. A., Unraveling 
         the mystery of Indian monsoon failure during El Ni\~no. 2006, {\it Science}, {\bf 314(5796)}, 115-119.}

\item {Pai, D. S., Latha, S., Rajeevan, M., Sreejith, O. P., Satbhai, N. S. and Mukhopadhyay, B., 
         Development of a new high spatial resolution (0.25$^{\circ} \times$ 0.25$^{\circ}$) long 
         period (1901-2010) daily gridded rainfall data set over India and its comparison with existing 
         data sets over the region. 2014, {\it Mausam}, {\bf 65}, 1-18.}  
         
\item {Pattanaik, D. R.,  Mukhopadhyay, B. and Kumar, A., Monthly Forecast of Indian 
         Southwest Monsoon Rainfall Based on NCEP’s Coupled Forecast System.  2012, 
         {\it Atm. Clim. Sci.}, {\bf 2(4)}, 479-491.}

\item {Pillai, P. A., Rao, S. A. and co-authors, Seasonal prediction skill of Indian summer 
         monsoon rainfall in NMME models and monsoon mission CFSv2. 
         2018, {\it Int. J. Climate}, {\bf 38}, e847-e861.}       
         
\item {Pokhrel, S., Saha, S. K. and co-authors, Seasonal prediction of Indian summer monsoon 
         rainfall in NCEP CFSv2: forecast and predictability error. 2016,  
         {\it Clim. Dyn.}, {\bf 46}, 2305-2326.} 
         
\item {Rajendran, K., Nanjundiah, R. S., Gadgil, Sulochana, and Srinivasan, J., How good 
         are the simulations of tropical SST-rainfall relationship by IPCC AR4 atmospheric 
         and coupled models?. 2012, {\it J. Earth Syst. Sci.}, {\bf 121(3)}, 595-610.}         
                          
\item {Ramu, D. A. and co-authors, Indian summer monsoon rainfall simulation and 
         prediction skill in the CFSv2 coupled model: Impact of atmospheric horizontal 
         resolution. 2016, {\it J. Geophys. Res. Atmos.}, {\bf 121(5)}, 2205-2221.}            
         
\item {Rao, S. A. and co-authors, MONSOON MISSION: A targeted activity to improve monsoon 
         prediction across scales. 2019, {\it Bull. Amer. Meteorol. Soc.}, {\bf 100(12)}, 2509-2532.}            
         
\item {Rasmusson, E. M. and Carpenter, T. H., The relationship between eastern equatorial 
         Pacific sea surface temperatures and rainfall over India and Sri Lanka. 1983,
         {\it Mon. Wea. Rev.}, {\bf 111(3)}, 517-528.}
            
\item {Rayner, N. A., Parker, D. A. and co-authors, Global analyses of SST, sea ice and night 
       marine air temperature since the late nineteenth century. 
       2003, {\it J. Geophys. Res. Atm.}, {\bf 108}, 4407, 10.1029/2002JD002670.}       

\item {Reynolds, R. W., Smith, T. M., Liu, C., Chelton, D. B., Casey, K. S. and Schlax, M. G., 
         Daily High-Resolution-Blended Analyses for Sea Surface Temperature. 
         2007, {\it J. Climate}, {\bf 20}, 5473-5496.}
                 
\item {Saha, S., Moorthi, S., and co-authors, The NCEP Climate Forecast System reanalysis.
         {\it Bull. Amer. Meteorol. Soc.}, 2010, {\bf 91}, 1015-1057.}
         
\item {Saha, S.,  Moorthi, S., Wu, X. and co-authors, The NCEP Climate Forecast System 
         version 2.  2014, {\it J. Climate}, {\bf 27}, 2185-2208.}
         
\item {Sajani, Surendran, Gadgil, Sulochana, Francis, P. A. and Rajeevan, M., Prediction of Indian rainfall 
         during the summer    
         monsoon season on the basis of links with equatorial Pacific and Indian Ocean climate indices.
         2015, {\it Environ. Res. Lett.}, {\bf 10},  094004.}
         
\item {Saji, N. H., Goswami, B. N., Vinayachandran, P. N., and Yamagata, T., A dipole in the 
         tropical Indian Ocean. 1999, {\it Nature}, {\bf 401}, 360-363.}
         
\item {Schaeybroeck, B. V. and Vannitsem, S., Postprocessing of Long-Range Forecasts. 
        2019, {\it In:  Statistical postprocessing of ensemble forecasts}, {\bf Chapter 10}, 267-290.}      

\item {Shukla, J. and Wallace, J. M., Numerical simulation of the atmospheric response to 
         equatorial Pacific sea surface temperature anomalies. 1983,
         {\it J. Atmos. Sci.}, {\bf 40}, 1613-1630.}

\item {Sikka, D. R., Some aspects of the large-scale fluctuations of summer monsoon rainfall 
          over India in relation to fluctuations in the planetary and regional scale circulation 
          parameters. 1980,  {\it Proc. Indian Acad. Sci. (Earth Planet. Sci.)}, 
         {\bf 89}, 179-195.}
        
 \item {Slingo, J. and Palmer, T. N., Uncertainty in weather and climate prediction. 2011,
         {\it Philos. Trans. R. Soc.}, {\bf 369}, 4751-4767.}   
         
\item {Sperber, K. R., and Palmer, T. N., Interannual Tropical Rainfall Variability in General 
         Circulation Model Simulations Associated with the Atmospheric Model Intercomparison 
         Project. 1996,  {\it J. Climate}, {\bf 9}, 2727-2750.}     
         
 \item {Torrence, C., and Webster, P. J., Interdecadal changes in the ENSO-monsoon system. 
          1999, {\it J. Climate}, {\bf 12(8)}, 2679-2690.}
                  
\item {Vishnu, S., Francis, P. A., Ramakrishna, S. S. V. S., and Schenoi, S. S. C., On the 
         relationship between the Indian summer monsoon rainfall and the EQUINOO in the 
         CFSv2. 2019,  {\it Clim. Dyn.}, {\bf 52}, 1263-1281.}
        
\item {Walker, G. T., and Bliss, E. W., World weather. V. 1932,  {\it Mem. Roy. Meteorol. Soc.}, 
         {\bf 4}, 53-84.}
  
 \item {Walker, G. T., Seasonal weather and its prediction. 1933,  {\it Nature}, 
         {\bf 132(3343)}, 805-808.}
         
\item {Webster, P. J. and Yang, S., Monsoon and ENSO: selectively interactive systems.
         1992,  {\it Quart. J. Roy. Meteorol. Soc.}, {\bf 118}, 877-926.}
               
\end{enumerate}
%
%
\newpage
\noindent\textbf{Appendix 1: Forecast Skill Scores}
\begin{justify}
The methods used for verifying forecasts are the following. 
\end{justify}
\begin{enumerate}
        \item Mean error is average error.
\begin{justify}
\[ Mean \,\, Error =\frac{1}{N} \sum_{i=1}^{N} \left( F_{i}-O_{i}\right)   \] 
\end{justify}
        \item Bias is comparison of average forecast magnitude to the observed.
\begin{justify}
 \[ BIAS=\frac{\frac{1}{N} \sum _{i=1}^{N}F_{i}}{\frac{1}{N} \sum_{i=1}^{N}O_{i}} \] 
\end{justify}
        \item RMSE is average magnitude of forecast errors and Anomaly correlation is 
                 comparison of forecast anomalies to observed.
\begin{justify}
\[ RMSE=\sqrt[]{\frac{1}{N} \sum _{i=1}^{N} \left( F_{i}-O_{i} \right) ^{2}} \] 
\end{justify}
        \item In addition, the amount of climatological JJAS rainfall over the Indian land region 
                 ($\mu$), the corresponding standard deviation of JJAS mean rainfall ($\sigma$) 
                 and its temporal coefficient of variation (CV) in percentage for 1982-2010, 
                 are estimated as:
\begin{justify}
\[  \mu =\frac{ \sum _{i=1}^{N}\text{India Rain}_{ \left( JJAS \right) _{i}}}{N} \] 
\end{justify}
\begin{justify}
 \[  \sigma =\sqrt[]{\frac{ \sum _{i=1}^{N} \left( \text{India Rain}_{ \left( JJAS \right) _{i}}- \mu  \right) ^{2}}{N}} \] 
\end{justify}
\begin{justify}
 \[  \]  \[ CV=\frac{ \sigma }{ \mu } \times 100 \] 
\end{justify}

\setlength{\parskip}{7.8pt}

\end{enumerate}

%
\newpage
\begin{table}
\large
\begingroup
\renewcommand{\arraystretch}{1} 
\begin{tabular}{|l|r|r|||r|r|r|r|r|}
\cline{1-8}
\textbf{Std. Anomalies}&\textbf{1983}&\textbf{1994}&\textbf{1985}&\textbf{1990}&\textbf{1997}& \textbf{1998}&\textbf{2006}\\
\toprule[1pt]
\cline{1-8}
\multirow{2}{*}{\textbf{IMD/HadISST}} & \cellcolor{lightseagreen}1.61 &   \cellcolor{lightseagreen}1.98 & \cellcolor{lightcoral}-0.89 &  0.98 & \cellcolor{lightseagreen}0.08 &  \cellcolor{lightcoral}0.04 &  \cellcolor{lightseagreen}1.22 \\\cline{2-8}
& \cellcolor{lightseagreen}-0.04&   \cellcolor{lightseagreen}0.46 & \cellcolor{lightcoral}-0.87 &  0.14 &  \cellcolor{lightseagreen}2.44 &  \cellcolor{lightcoral}-1.00 &  \cellcolor{lightseagreen}0.39 \\\hline
\multirow{2}{*}{\textbf{January ICs}} & \cellcolor{lightcoral}0.47 &  0.19 &  0.31 & \cellcolor{lightseagreen}-1.17 & \cellcolor{lightcoral}-0.80 &  \cellcolor{lightcoral}2.76 & \cellcolor{lightseagreen}-0.51 \\
&\cellcolor{lightcoral} -0.96 & -0.42 & -1.22 &  \cellcolor{lightseagreen}0.92 & \cellcolor{lightcoral}0.71 &  \cellcolor{lightcoral}-2.16 &  \cellcolor{lightseagreen}0.32\\ \hline
\multirow{2}{*}{\textbf{February ICs}} &  \cellcolor{lightcoral}1.20 &  \cellcolor{lightcoral}0.86 &  \cellcolor{lightcoral}0.34 & \cellcolor{lightseagreen}-1.25 & \cellcolor{lightseagreen}-1.50 &  \cellcolor{lightcoral}2.01 & \cellcolor{lightseagreen}-0.99 \\
 & \cellcolor{lightcoral}-1.11 & \cellcolor{lightcoral}-1.13 &  \cellcolor{lightcoral}-1.12 &  \cellcolor{lightseagreen}1.58 & \cellcolor{lightseagreen}0.89 &  \cellcolor{lightcoral}-1.80 & \cellcolor{lightseagreen}1.19 \\\hline
\multirow{2}{*}{\textbf{March ICs}}   & \cellcolor{lightcoral} 0.71 & \cellcolor{lightcoral}-0.40 &  \cellcolor{lightcoral}0.98 & -0.73 & \cellcolor{lightseagreen}-1.03 &  \cellcolor{lightcoral}1.25 & \cellcolor{lightseagreen}-1.08 \\
 & \cellcolor{lightcoral}-0.52 & \cellcolor{lightcoral}-0.40 &  \cellcolor{lightcoral}-1.17 &  0.67 & \cellcolor{lightseagreen}1.02 &  \cellcolor{lightcoral}-1.23 & \cellcolor{lightseagreen}1.09 \\\hline
\multirow{2}{*}{\textbf{April ICs}}     & -0.06 &  0.13 &  \cellcolor{lightcoral}0.93 & -0.36 & \cellcolor{lightseagreen}-1.89 &  \cellcolor{lightcoral}2.03 & \cellcolor{lightseagreen}-0.63 \\
 &  0.04 & -0.23 &  \cellcolor{lightcoral}-1.20 & -0.01 & \cellcolor{lightseagreen}1.81 &  \cellcolor{lightcoral}-1.19 & 0.\cellcolor{lightseagreen}65 \\\hline
\multirow{2}{*}{\textbf{May ICs}}     & -1.11 & \cellcolor{lightseagreen}-0.14 &  \cellcolor{lightcoral}1.38 & \cellcolor{lightcoral}-0.75 & \cellcolor{lightseagreen}-2.07 &  \cellcolor{lightcoral}1.81 & \cellcolor{lightseagreen}-0.85\\
    &  0.59 &  \cellcolor{lightseagreen}0.10 &  \cellcolor{lightcoral}-1.13 &  \cellcolor{lightcoral}0.37 & \cellcolor{lightseagreen}2.32 &  \cellcolor{lightcoral}-2.05 & \cellcolor{lightseagreen}0.92\\\hline
\bottomrule[1pt]
\end{tabular}
\endgroup
\caption{Standardised anomalies of ISMR and ENSO index for special years of 1983 and 1994, 
              and 1985, 1990, 1997, 1998 and 2006, for ensemble means of CFSv2-NCEP reforecasts 
              with January (L0) to May (L4) initial conditions (ICs). For each entity, the first and the
              second sub-rows are the anomalies of ISMR and ENSO index respectively. 
              Cells are coloured in green (red) if anomaly of EQUINOO index is positive (negative) for each year.}
\end{table}

\newpage
\begin{table}
\large
\begingroup
\setlength{\tabcolsep}{4pt} 
\renewcommand{\arraystretch}{1} 
\begin{tabular}{|l|r|r|r|}
\cline{1-4} 
\textbf{IAV}  & \textbf{CFSv2-NCEP}  & \textbf{CFSv2-CSIR}  & \textbf{Observation} \\
\textbf{Skill Scores} & \textbf{Feb. ICs. Ens.} &  \textbf{Apr/May Ens.} & \textbf{(IMD)} \\
 \cline{1-4}
Mean Error (IAV) & {\color{blue} -3.11} & {\color{blue} -1.99} & \\
Bias (IAV)& {\color{blue} 0.64} & {\color{blue} 0.69} & \\
RMSE (IAV)& {\color{blue} 3.14} &  {\color{blue} 2.12} & \\ 
ISMR $\mu$ (mm/day) &  3.50 &   4.51 & 6.50 \\
ISMR $\sigma$ (mm/day) &  0.51 &  0.66 & 0.76 \\
ISMR  CV (\%) &  14.5 &  10.3 &11.7\\ \cline{1-4}
\end{tabular}
\endgroup
\vspace{0.25 cm}
\caption{Skill scores for interannual variation (IAV) of standardized (with standard deviation) 
              anomalies of Indian summer monsoon rainfall averaged over monsoon region (ISMR), 
              for ensemble means of CFSv2-NCEP reforecasts with February (L3) initial conditions (ICs) 
              and CFSv2-CSIR reforecasts with late-April/early-May ICs, with respect to corresponding 
              IMD observation for 1982-2010. Time series statistics against the observed are mean error, 
              bias, and RMSE are shown as skill scores. The mean ($\mu$), standard deviation 
              ($\sigma$) and coefficient of variation (CV in \%) of ISMR for models and IMD 
              observation are also given.}
\end{table}

\newpage
\begin{table}
{
\large
\begin{tabular}{@{}|l|l|r|r|r|@{}}
\cline{1-5}
\multicolumn{1}{|c|}{\textbf{CFSv2 reforecasts}} & \multicolumn{1}{c|}{\textbf{ICs}} & \multicolumn{1}{c|}{\textbf{PCC}} & \multicolumn{1}{c|}{\textbf{SD}} & \multicolumn{1}{c|}{\textbf{bias}} \\ \cline{1-5} 
\multirow{5}{*}{\begin{tabular}[c]{@{}l@{}} Ensemble means of\\ CFSv2-NCEP reforecasts with L4-L0 ICs\end{tabular}}   & Jan (L4) & 0.69 & 0.86 & \multicolumn{1}{r|}{-33.2\%} \\ \cline{2-5} 
 & Feb (L3)  & 0.69 & 0.85 & \multicolumn{1}{r|}{ \,\,-34.0\%}  \\ \cline{2-5} 
 & Mar (L2)  & 0.69 & 0.86 & \multicolumn{1}{r|}{ -33.2\%} \\ \cline{2-5} 
 & Apr (L1)  & 0.69 & 0.90 & \multicolumn{1}{r|}{-27.6\%} \\ \cline{2-5} 
 & May (L0)  & 0.71 & 0.96 & \multicolumn{1}{r|}{-19.6\%} \\ \cline{1-5} 
\multirow{6}{*}{\begin{tabular}[c]{@{}l@{}} CFSv2-CSIR reforecasts and \\ ensemble mean \end{tabular}} & 21 Apr & 0.69 & 0.92 & \multicolumn{1}{r|}{-27.4\%} \\ \cline{2-5} 
 & 26 Apr & 0.69 & 0.95 & \multicolumn{1}{r|}{-24.3\%} \\ \cline{2-5} 
 & 01 May & 0.70 & 0.93 & \multicolumn{1}{r|}{-24.3\%} \\ \cline{2-5} 
 & 06 May & 0.71 & 0.95 & \multicolumn{1}{r|}{-21.6\%} \\ \cline{2-5} 
 & 11 May & 0.70 & 0.95 & \multicolumn{1}{r|}{-19.8\%} \\ \cline{2-5} 
 & 5 ICs Mean & 0.70 & 0.94 & \multicolumn{1}{r|}{-23.5\%} \\ \cline{1-5} 
\end{tabular}
\vspace{0.25 cm}
\caption{Statistics for climatological summer (JJAS) mean rainfall over India for ensemble 
              means of CFSv2-NCEP reforecasts with January (L0) to May (L4) initial conditions 
              (ICs) and CFSv2-CSIR reforecasts with late-April/early-May ICs and their ensemble 
              mean with respect to corresponding IMD observation for 1982-2010 period. Statistics 
              are spatial pattern correlation coefficient (PCC), ratio of model spatial standard 
              deviation with the observed (SD) and bias in simulating the mean with respect to 
              IMD observation.}
}
\end{table}

\newpage
\begin{table}
\begingroup
\begin{tabular}{|c|c|c|}
\cline{1-3}
\multicolumn{3}{|c|}{\textbf{Cold Events}}\\\cline{1-3}
\cline{1-3}
\textbf{HadISST}&\textbf{CFSv2-NCEP}&\textbf{CFSv2-CSIR}\\
\toprule[1pt]
\cline{1-3}
& \cellcolor{lightseagreen}1983 &  \\\cline{1-3}
1984 & 1984 & 1984 \\
1985 & 1985 & 1985 \\
&   & \cellcolor{lightseagreen}1986 \\
1988 & 1988 & 1988 \\
1989 & \cellcolor{lightcoral} & 1989 \\
& \cellcolor{lightseagreen} 1994 & \\
& \cellcolor{lightseagreen} 1995 & \cellcolor{lightseagreen} 1995 \\
1998 & 1998 & 1998 \\
1999 & \cellcolor{lightcoral} & 1999 \\
2000 & \cellcolor{lightcoral} & \cellcolor{lightcoral} \\
&  & \cellcolor{lightseagreen} 2003 \\
2007 & 2007 & 2007 \\
2010 & \cellcolor{lightcoral} & 2010 \\\hline
\bottomrule[1pt]
\end{tabular}
\begin{tabular}{|c|c|c||}
\cline{1-3}
\multicolumn{3}{|c|}{\textbf{Warm Events}}\\\cline{1-3}
\cline{1-3}
\textbf{HadISST}&\textbf{CFSv2-NCEP}&\textbf{CFSv2-CSIR}\\
\toprule[1pt]
\cline{1-3}
1982 & \cellcolor{lightcoral} & \cellcolor{lightcoral}\\
1987 & 1987 & \cellcolor{lightcoral} \\
&   \cellcolor{lightseagreen}1990 & \\
1991 & \cellcolor{lightcoral} & \cellcolor{lightcoral} \\
&   \cellcolor{lightseagreen}1993 & \\
1997 & 1997 & 1997 \\
&   \cellcolor{lightseagreen}2001 &  \cellcolor{lightseagreen}2001\\
2002 & 2002 & 2002 \\
2004 & 2004 & 2004 \\
&   \cellcolor{lightseagreen}2006 &  \cellcolor{lightseagreen}2006\\
2009 & 2009 & 2009 \\\hline
\bottomrule[1pt]
\end{tabular}
\endgroup
\caption{List of cold events (left panel) and warm events (right panel) over NINO3.4 from observation  (HadISST) and CFSv2-NCEP reforecasts with February (L3) ICs and CFSv2-CSIR reforecasts with late-April/early-May ICs. Cold and warm events are defined as the standardized Boreal summer (JJAS) SST anomaly being less than -0.5 and greater than 0.5 respectively. False alarms are highlighted in green and misses are coloured in red.}
\end{table}

\newpage 
\begin{figure}[ht]
\centering
\includegraphics[trim={1cm 5.5cm 1cm 4cm},clip, scale=0.64]{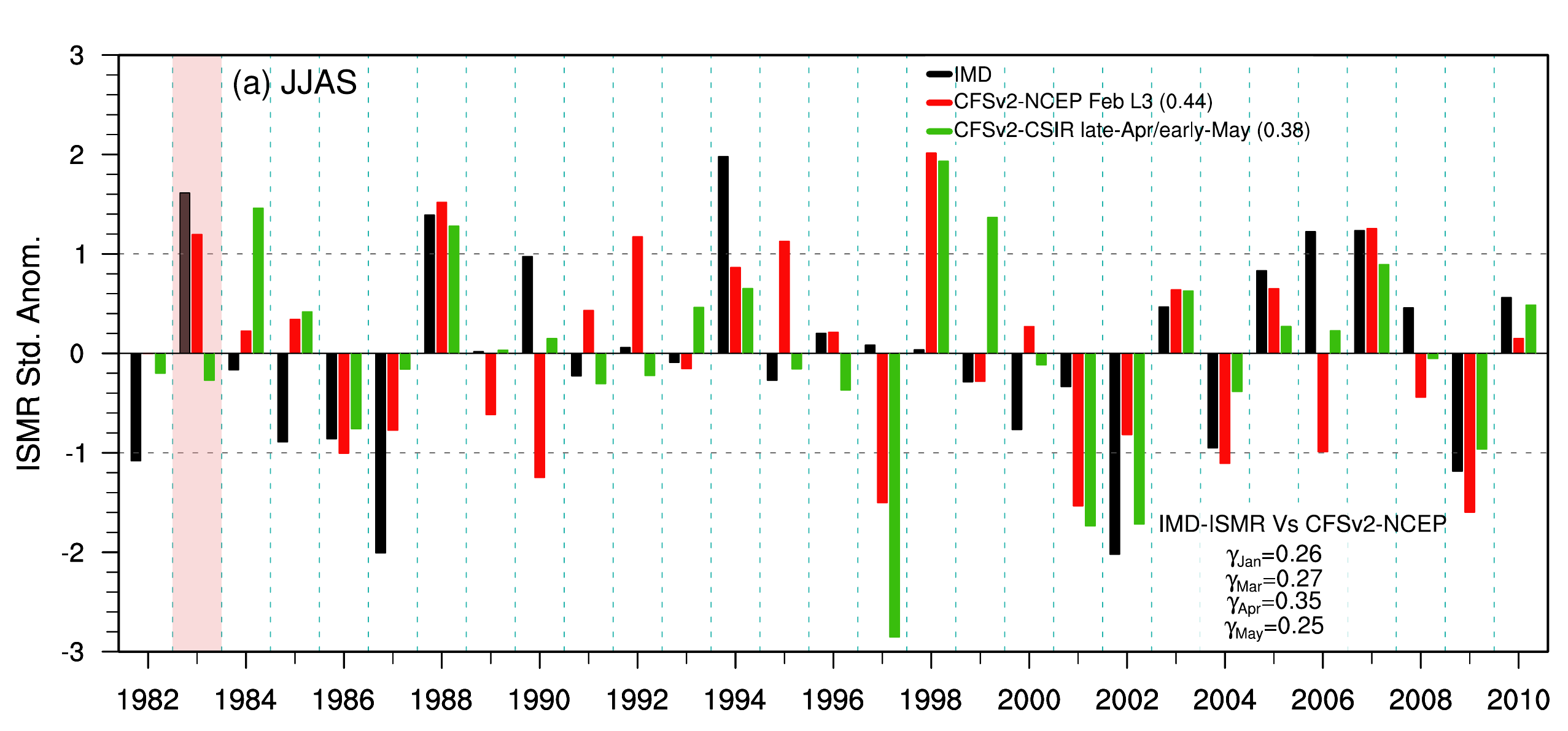}
\includegraphics[trim={1cm 5.5cm 1cm 4cm},clip, scale=0.64]{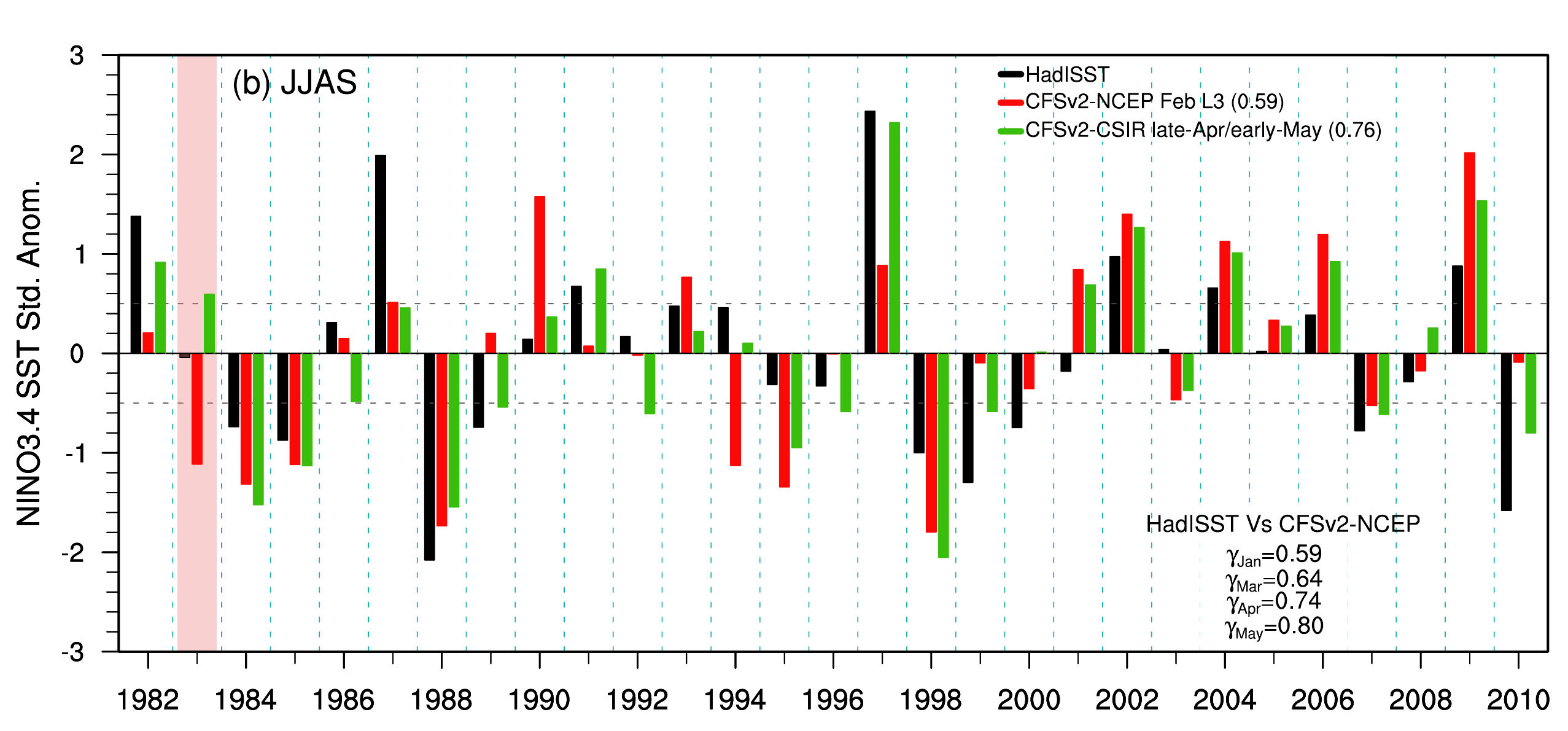}
 \caption{a) Interannual variability of standardised anomalies of Indian summer monsoon rainfall 
              over monsoon region (ISMR) in the ensemble means of CFSv2-NCEP reforecasts with 
              February (L3) initial conditions (ICs) and CFSv2-CSIR reforecasts with late-April/early-May 
              ICs, are compared with the corresponding ISMR anomalies from IMD observation.  
              Correlation coefficients ($\gamma$) between observed ISMR anomalies and those from the ensemble 
              means of CFSv2-NCEP reforecasts with L4 to L0, and CFSv2-CSIR reforecasts with 
              late-Apr/early-May ICs, are also given. 1983 ISMR anomalies are highlighted in light red 
              background color. b) Same as a) but for anomalies of ENSO index defined as standardised 
              anomalies of NINO3.4 SST. Correlation coefficients between observed HadISST based 
              ENSO index anomalies and those from the ensemble means of CFSv2-NCEP reforecasts 
              with L4 to L0, and CFSv2-CSIR reforecasts with late-Apr/early-May ICs, are also given.}
\end{figure}

\newpage 
\begin{figure}[ht]
 \centering
\includegraphics[clip, scale=1.1]{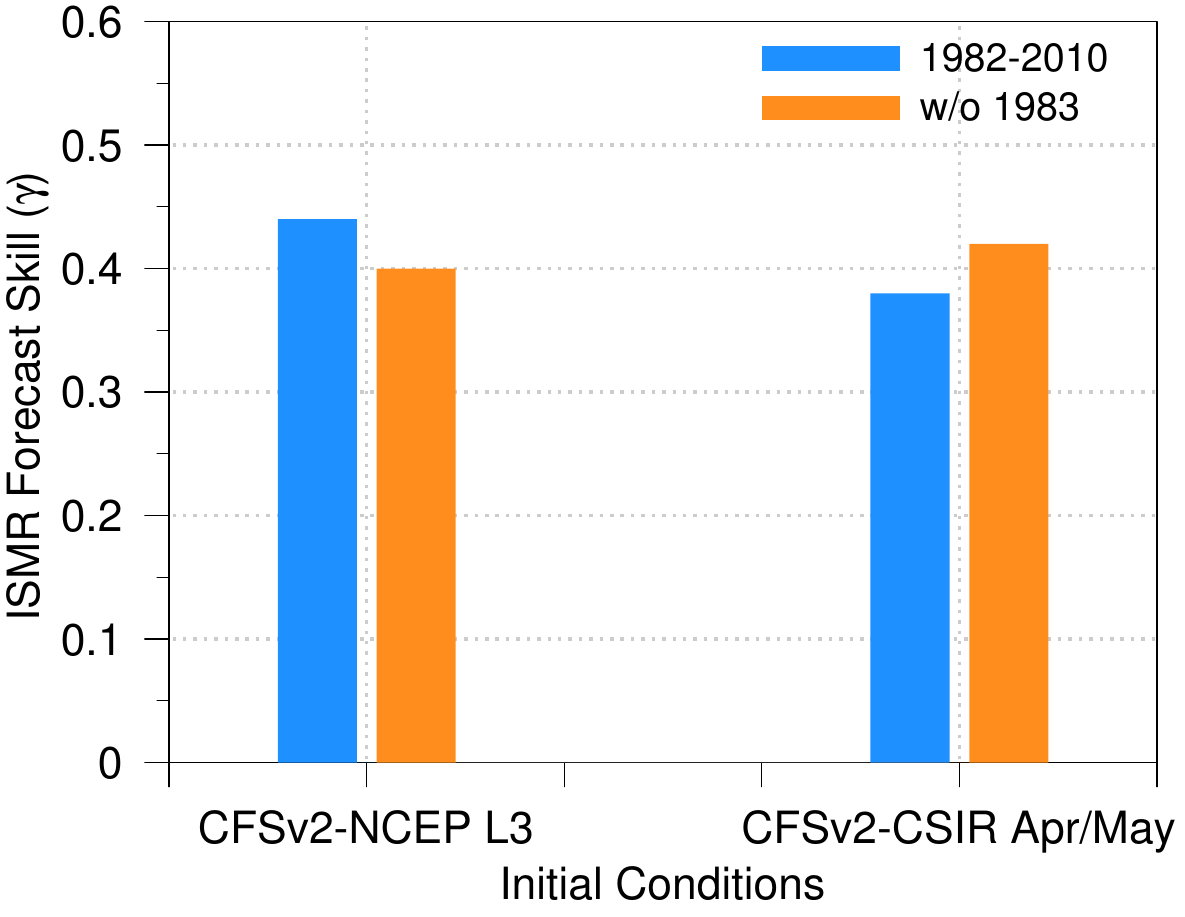}
\caption{Correlations of standardised ISMR anomalies of deterministic (ensemble mean) 
              CFSv2-NCEP reforecasts with February (L3) initial conditions (ICs) and CFSv2-CSIR 
              reforecasts with late-April/early-May ICs, against the IMD based observation for 
              i) 1982-2010 period and ii) 1982-2010 period excluding 1983.}
\end{figure}

\newpage 
\begin{figure}[ht]
 \centering
 \includegraphics[trim={7.5cm 0.5cm 7.5cm 0.5cm},clip,scale=1.2]{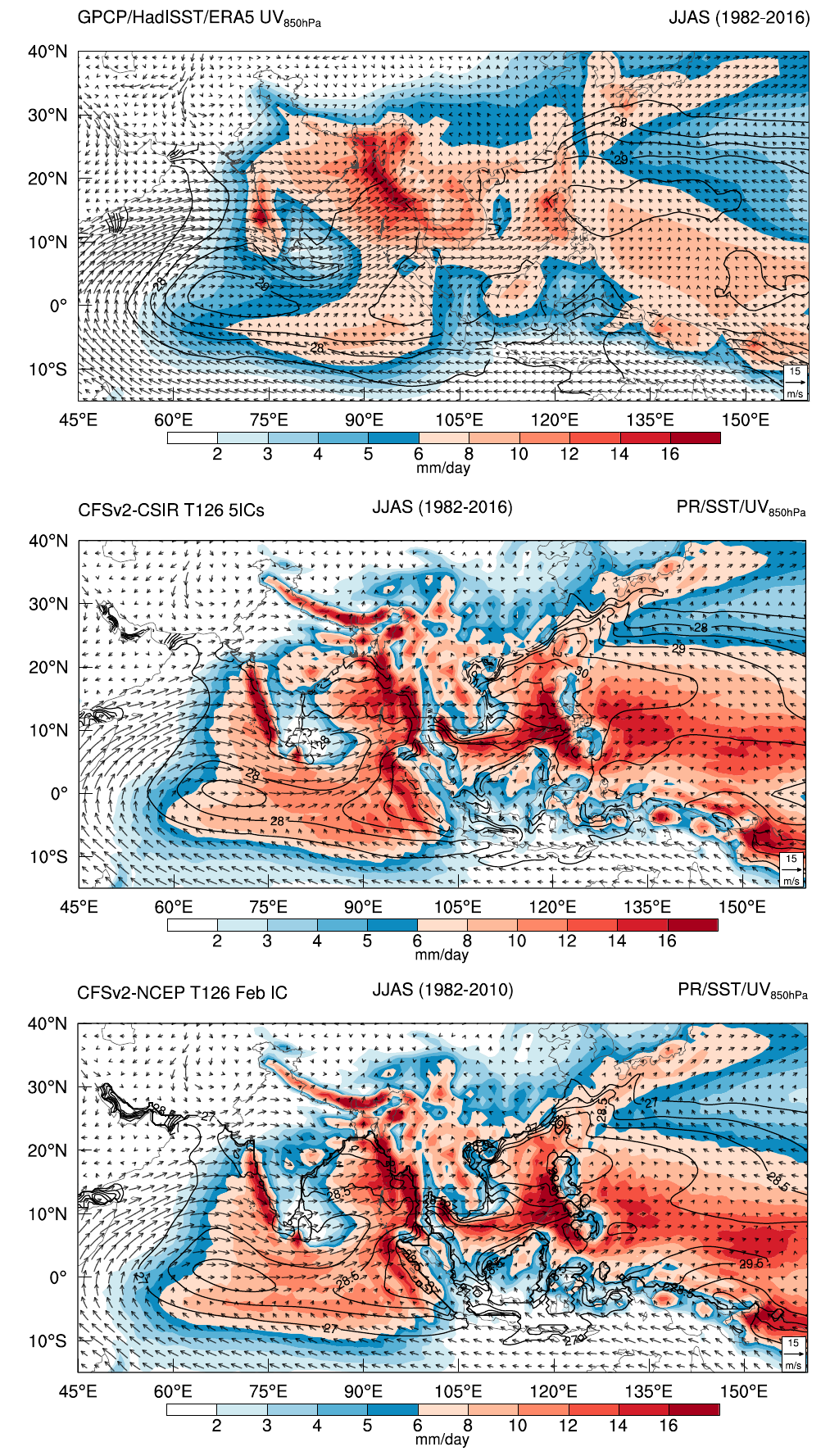}
 \caption{Climatological summer mean (JJAS) rainfall (shaded), SST (contours) and 850 hPa 
              winds (vectors) from i) observation (top), and ensemble means of ii) CFSv2-CSIR 
              reforecasts with late-April/early-May ICs (middle) and iii) CFSv2-NCEP reforecasts 
              with L3 ICs (bottom).}
\end{figure}

\newpage 
\begin{figure}[ht]
 \centering
 \includegraphics[trim={1.5cm 5cm 3cm 6cm},clip, scale=0.9]{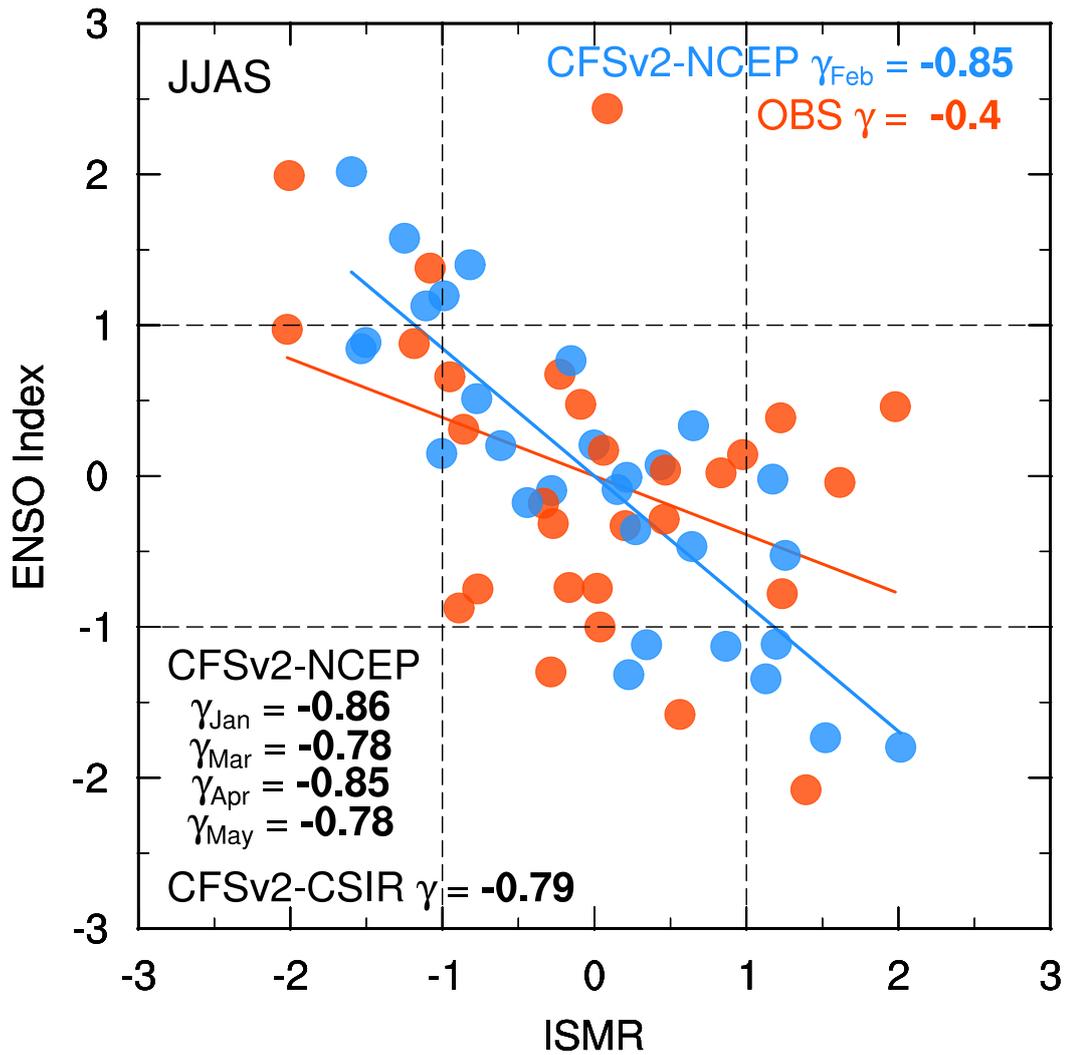}
 \caption{Anomalies of ISMR plotted against ENSO index for CFSv2-NCEP L3 (blue) and 
              observation (red). Respective regression lines are drawn. Correlations ($\gamma$) for ensemble 
              means of CFSv2-NCEP L4, L2, L1 and L0, and CFSv2-CSIR with late-Apr/early-May 
              ICs, are given in bottom-left corner and for CFSv2-NCEP L3 and observation are 
              given in top-right corner.}
\end{figure}

\newpage 
\begin{figure}[ht]
 \centering
 \includegraphics[trim={2.5cm 3.5cm 2.0cm 4.5cm},clip, scale=0.9]{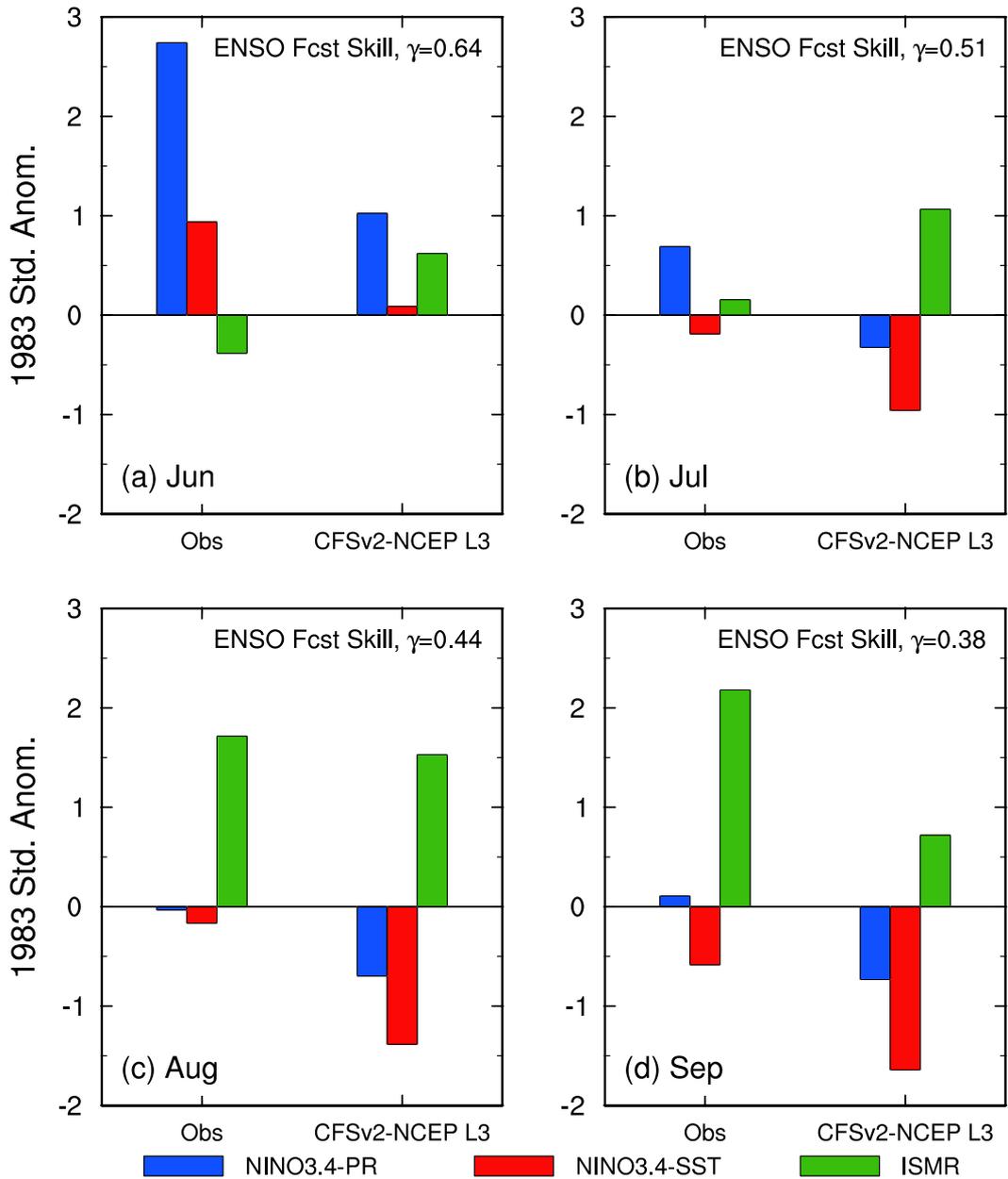}
 \caption{Monthly standardised anomalies of NINO3.4 rainfall (NINO3.4-PR) plotted along 
              with the corresponding anomalies of ENSO index (NINO3.4-SST) and ISMR from 
              observation and ensemble mean of CFSv2-NCEP L3 for a) June, b) July, 
              c) August and d) September. Corresponding correlations ($\gamma$) are also given.}
\end{figure}

\newpage 
\begin{figure}
 \centering
  \includegraphics[trim={0cm 0cm 0cm 0cm},clip, scale=1.4]{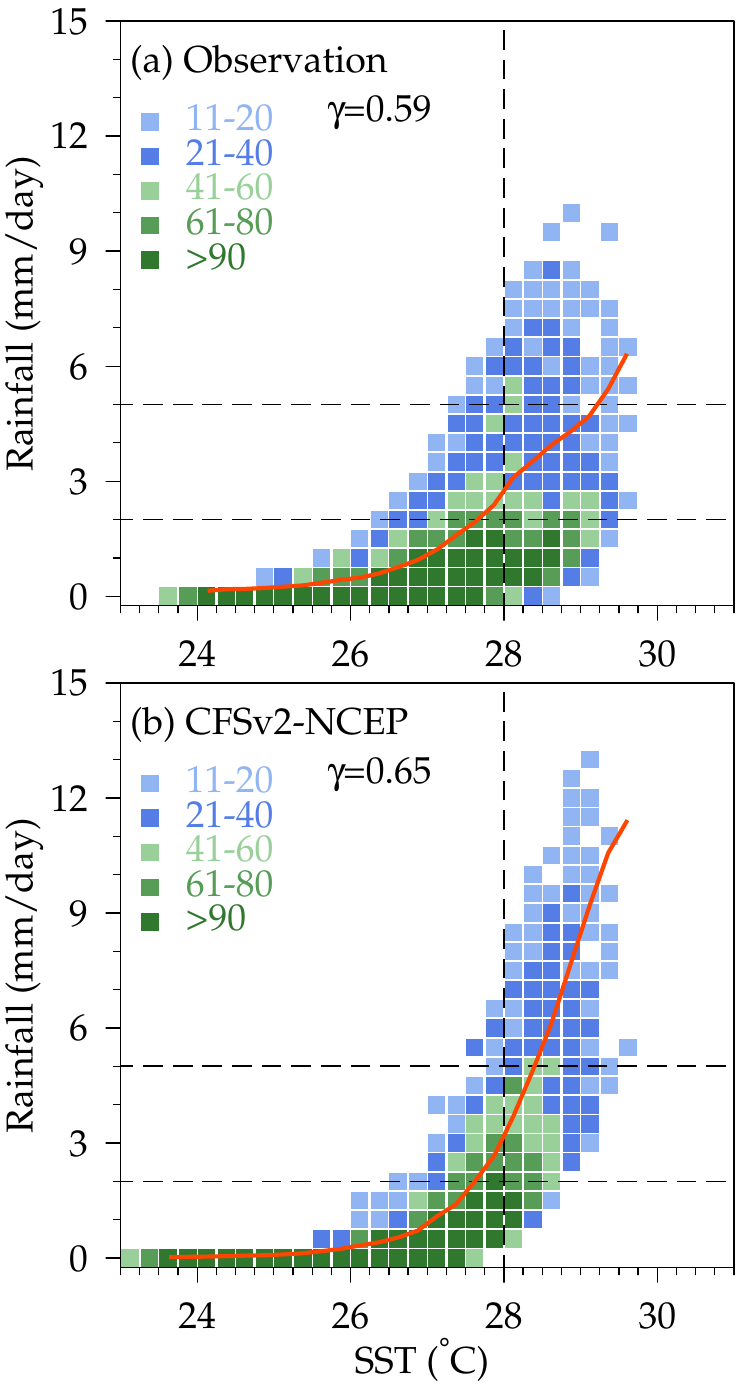}
  \caption{Distribution of the number of points for each 0.25$^{\circ}$C SST and 0.5 mm/day 
               rainfall bins along with the mean rainfall versus mean SST for each bin (red curve) 
               showing the relationship between rainfall and SST, for June, July, August and 
               September of 1982-2010 period over NINO3.4 in e) observation and f) ensemble 
               mean of CFSv2-NCEP L3. }
\end{figure}

\newpage 
\begin{figure}
 \centering
 \includegraphics[trim={1.5cm 4cm 1cm 3cm},clip, scale=0.85]{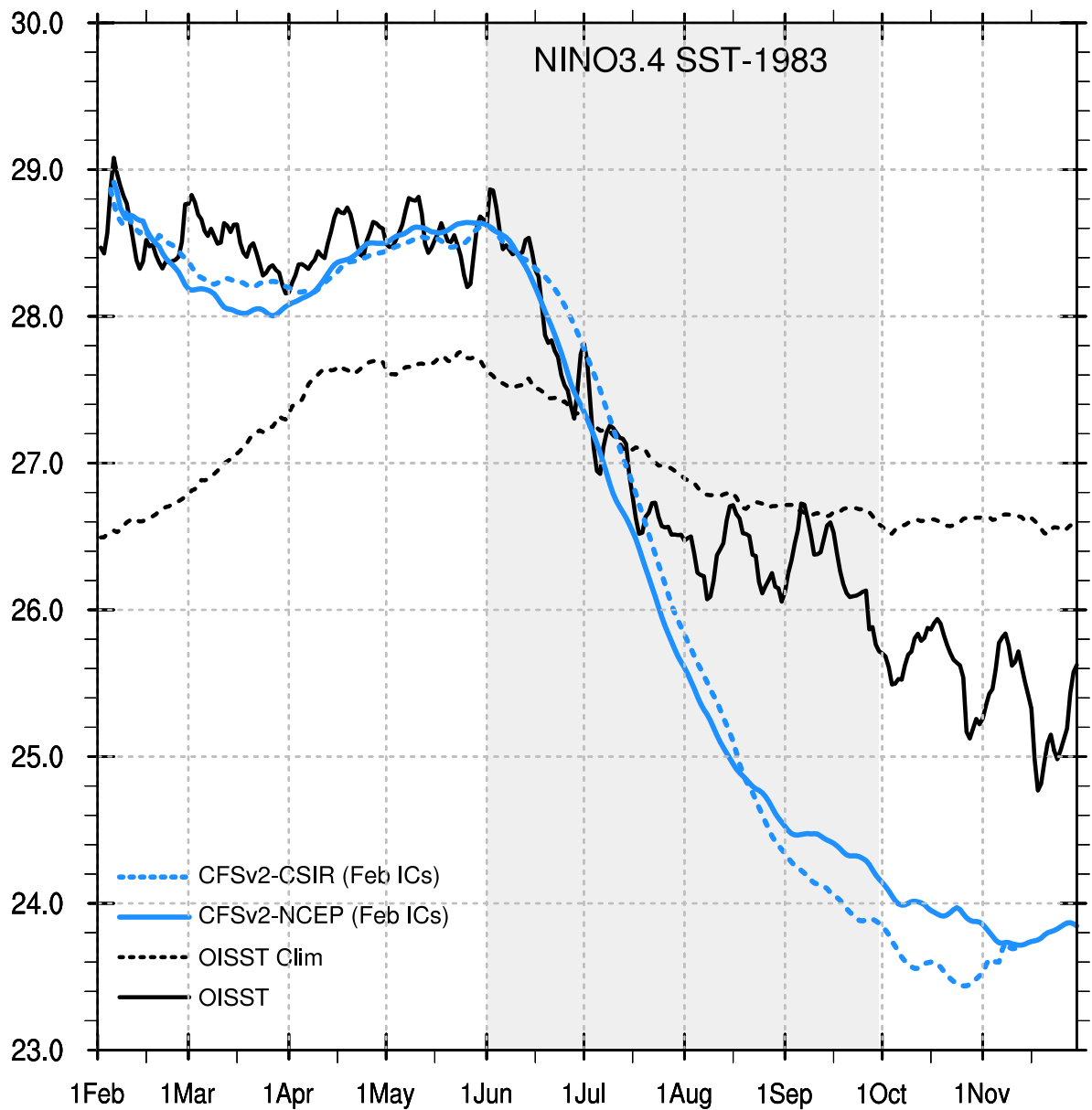}
 \caption{Evolution of daily SST averaged over NINO3.4 from OISST observation, 
              OISST daily climatology and ensemble mean CFSv2-NCEP reforecasts with 
              February (L3) initial conditions (ICs), and ensemble mean CFSv2-CSIR 
              reforecasts (current version of CFSv2 being used in India) with February 
              (L3) ICs.}
\end{figure}

\clearpage
\begin{figure}[bt]
\centering
\includegraphics[trim={1.5cm 6cm 1cm 5cm},clip, scale=0.6]{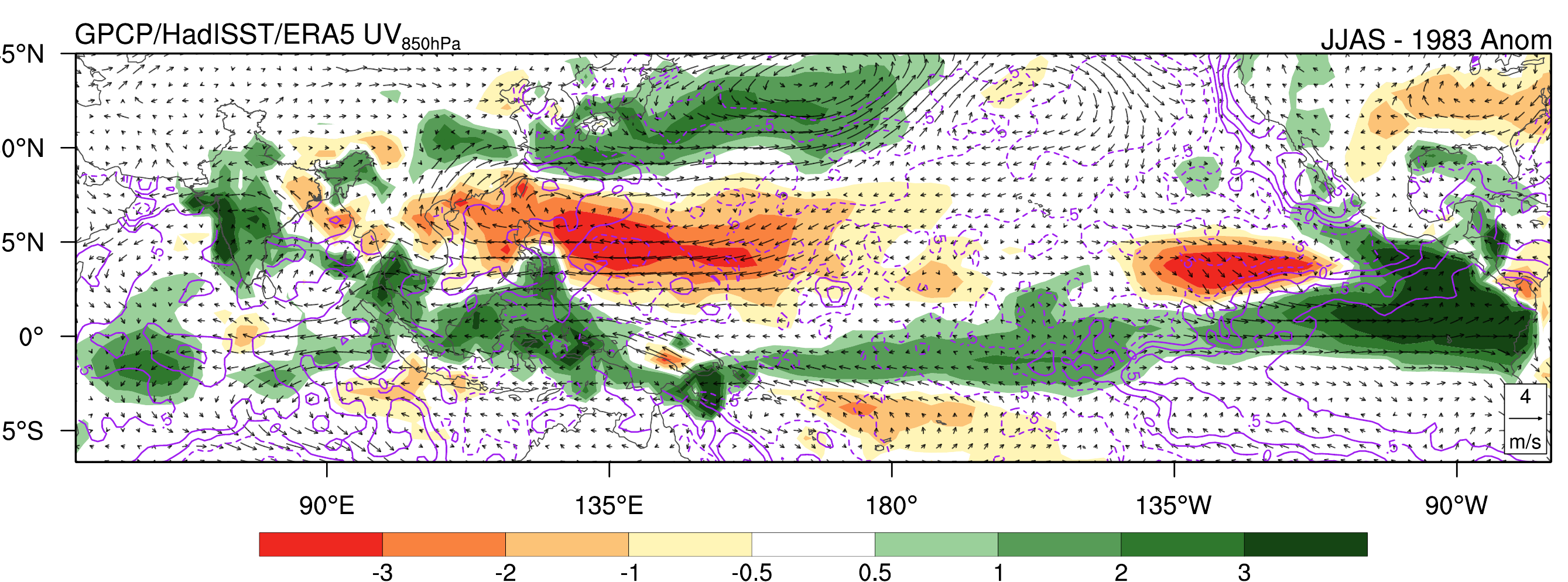}
\includegraphics[trim={1.5cm 6cm 1cm 5cm},clip, scale=0.6]{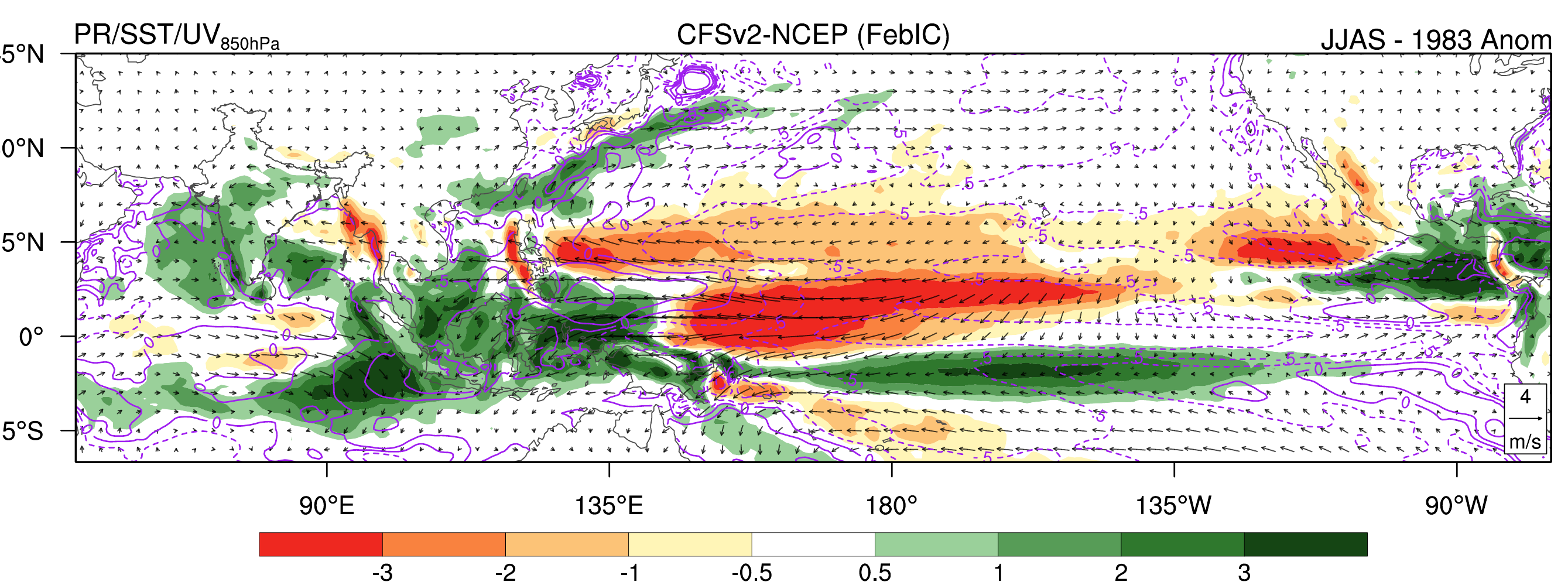}
\includegraphics[trim={1.5cm 6cm 1cm 5cm},clip, scale=0.6]{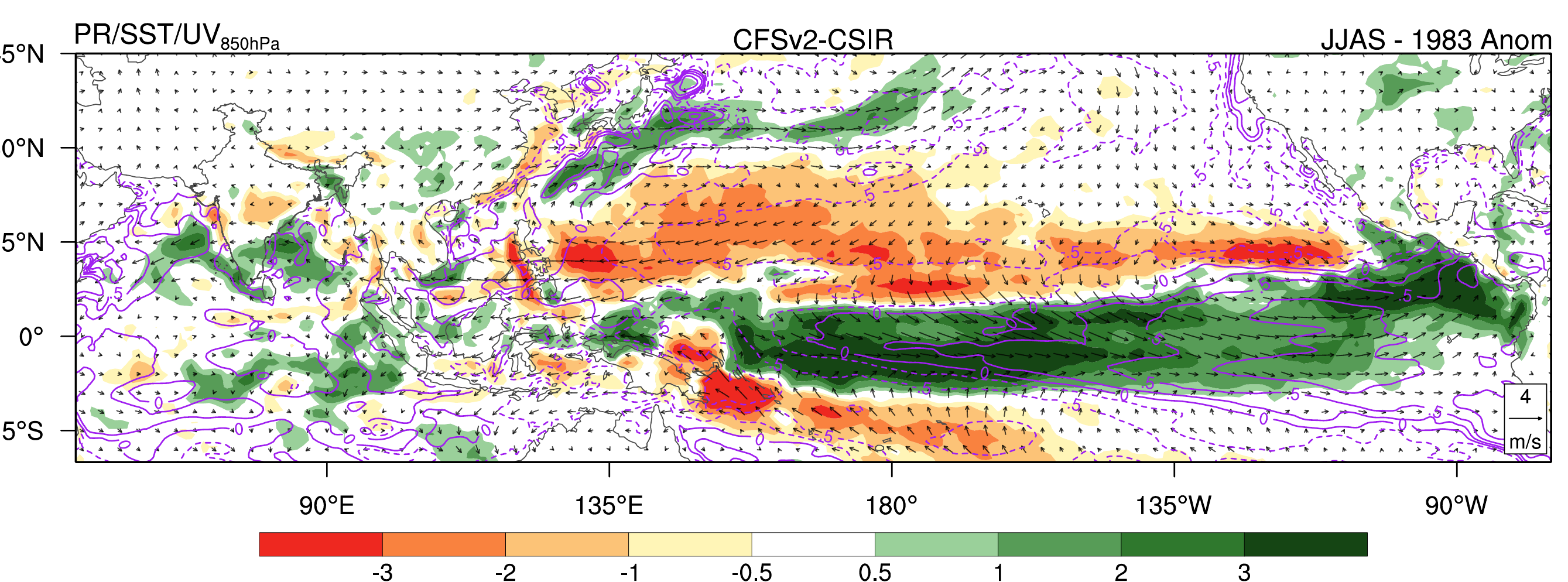}
\caption{Seasonal summer mean (JJAS) anomalies of rainfall (shaded), SST (contour) and 850 hPa winds (vectors) from i) observation (top), and ensemble means of ii) CFSv2-NCEP reforecasts with L3 ICs (middle) and iii) CFSv2-CSIR reforecasts with 5 ICs (bottom).}
\end{figure}

\begin{figure}
 \centering
 \includegraphics[trim={2cm 5.2cm 3.5cm 7.1cm},clip, scale=0.65]{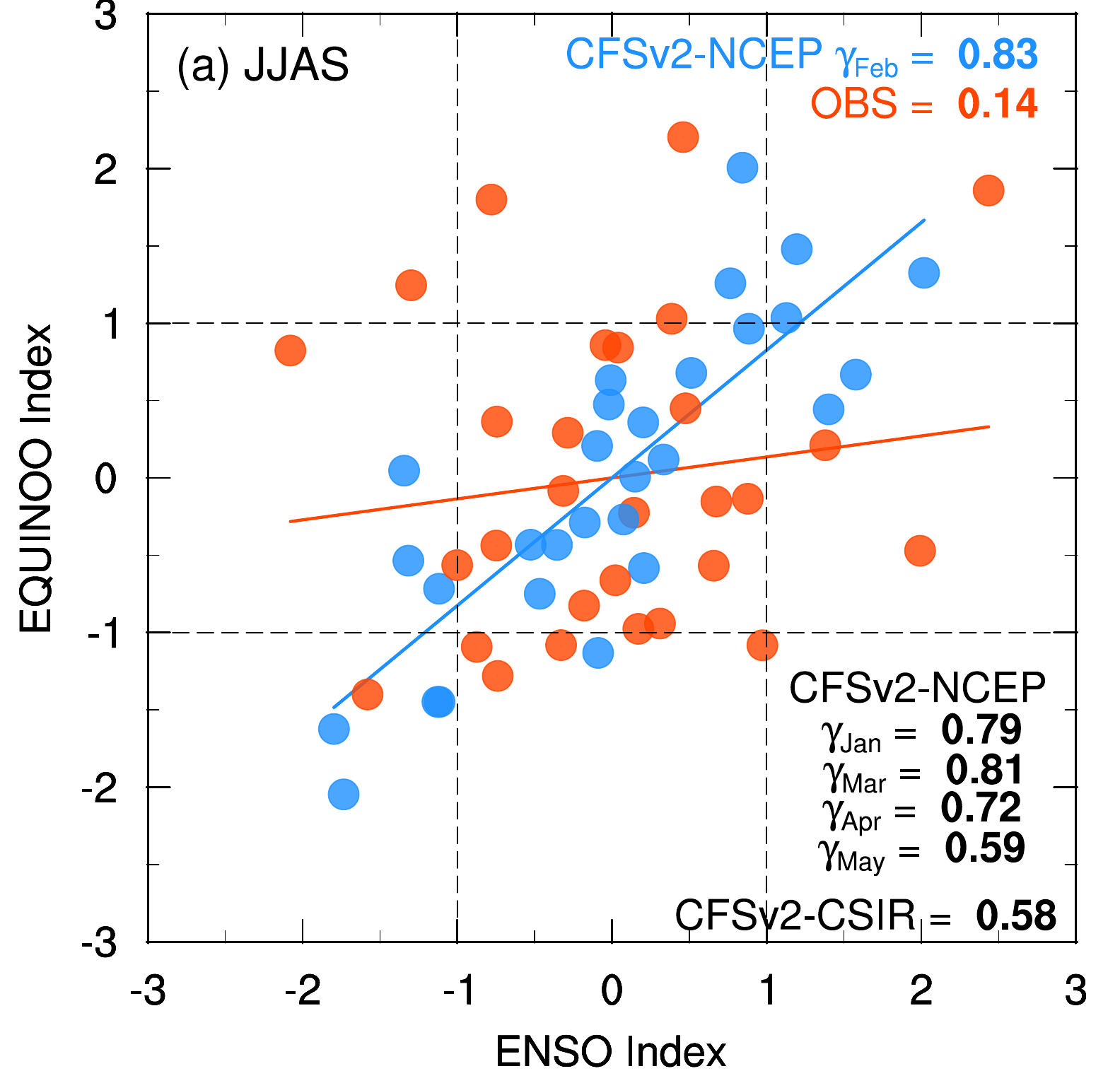}
 \includegraphics[trim={2cm 5.2cm 3.5cm 7.1cm},clip, scale=0.65]{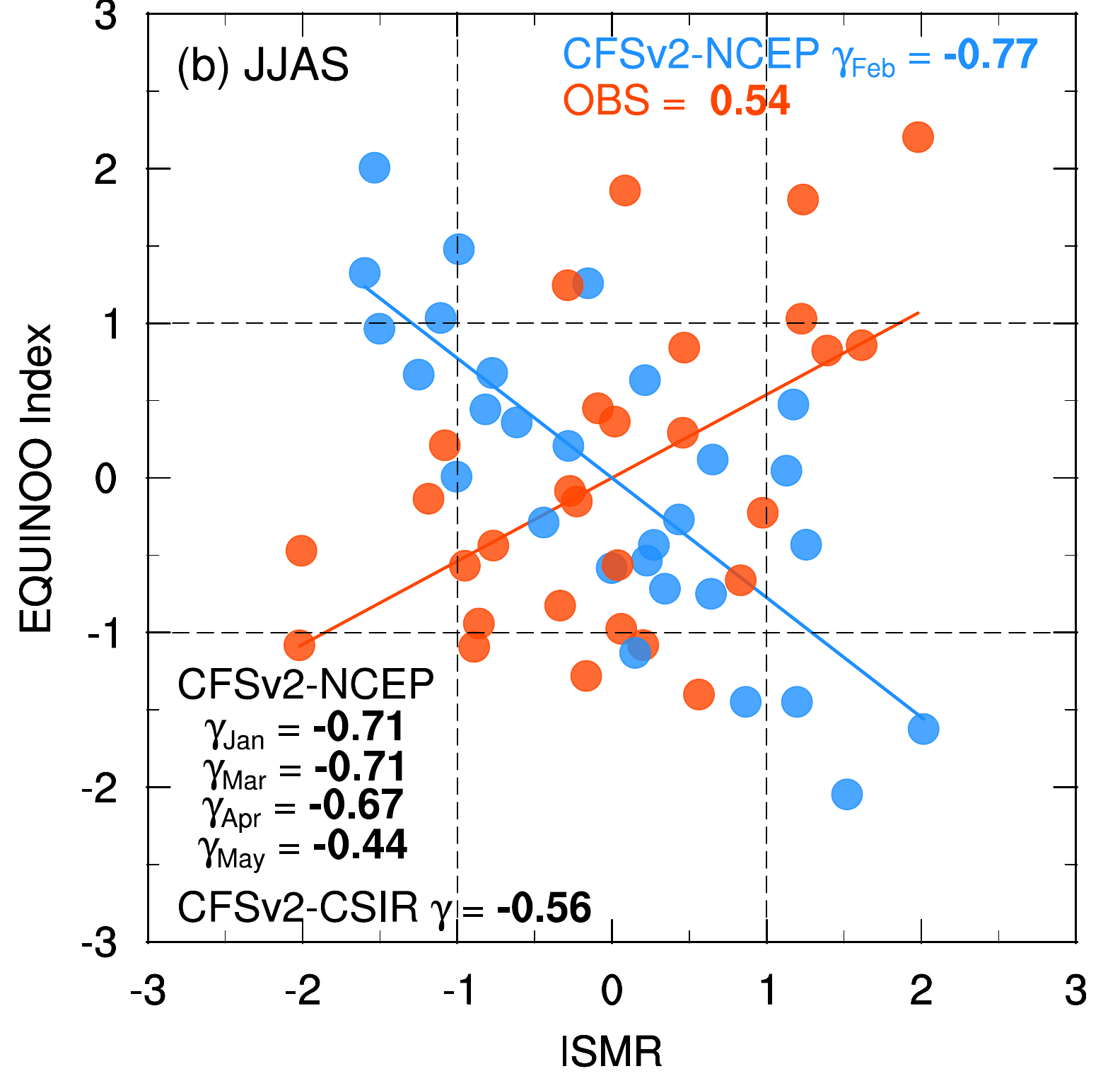}
 \caption{a) Anomalies of EQUINOO index plotted against ENSO index for 
              CFSv2-NCEP L3 (blue) and observation (red). b) Anomalies of EQUINOO index  
              plotted against ISMR for CFSv2-NCEP L3 (blue) and 
              observation (red). Respective regression lines are drawn and 
              correlations ($\gamma$) for CFSv2-NCEP L4, L2, L1 and L0, and CFSv2-CSIR 
              with late-April/early-May ICs, are given in bottom-left corner and for 
              CFSv2-NCEP L3 and observation are given in top-right corner of 
              a and b.}
\end{figure}

\end{document}